\documentclass[10pt,aps,prd,twocolumn,showpacs,superscriptaddress,eqsecnum,longbibliography,nofootinbib]{revtex4-1}
\usepackage{amsmath,amsfonts,amssymb,amsthm,mathrsfs,latexsym,epsfig,enumerate,times}

\usepackage{xcolor}
\definecolor{navy}{RGB}{0,0,150}
\usepackage[colorlinks, linkcolor=navy, citecolor=navy, urlcolor=navy, plainpages=false, pdfstartview=FitH]{hyperref}
\usepackage{appendix}
\allowdisplaybreaks
\newcommand{\GZU}{School of Physics, Guizhou University, Guiyang 550025, China}
\newcommand{\UOW}{Faculty of Physics, University of Warsaw, Pasteura 5, 02-093 Warsaw, Poland}
\newcommand{\UOE}{Institut f\"ur Quantengravitation, Friedrich-Alexander-Universit\"at Erlangen-N\"urnberg, Staudtstr. 7/B2, 91058 Erlangen, Germany}

\newcommand{\SCUT}{Department of Physics, South China University of Technology, Guangzhou 510641, China}

\newcommand{\sst}{\scriptscriptstyle}

\allowdisplaybreaks

\begin{document}

\title{Alternative $k=-1$ loop quantum cosmology}

\author{Jinsong Yang}
\email{jsyang@gzu.edu.cn}
\affiliation{\GZU}

\author{Cong Zhang}
\email{cong.zhang@gravity.fau.de}
\affiliation{\UOE}
\affiliation{\UOW}

\author{Xiangdong Zhang}
\thanks{Corresponding author}
\email{scxdzhang@scut.edu.cn}
\affiliation{\SCUT}

\begin{abstract}

An alternative quantization of the gravitational Hamiltonian constraint of the $k=-1$ Friedmann-Robertson-Walker model is proposed by treating the Euclidean term and the Lorentzian term independently, mimicking the treatment of full loop quantum gravity. The resulting Hamiltonian constraint operator for the $k=-1$  model with a massless scalar field is successfully constructed, and is shown to have the corrected classical limit. Compared to the former quantization schemes  in the literature where only the Euclidean term is quantized, the new quantum dynamics of the $k=-1$  model with a massless scalar field indicates that the classical big-bang singularity is replaced by an asymmetric quantum bounce.

\end{abstract}

\maketitle

\section{Introduction}\label{section1}

How to quantize general relativity (GR) in a consistent manner is a great challenge to theoretical physics. One of the promising candidates is the so-called loop quantum gravity (LQG) which is a nonperturbative approach to quantum GR \cite{Rovelli:2004tv,Thiemann:2007pyv,Ashtekar:2004eh,Han:2005km}. In the past three decades, LQG has made remarkable progress, such as making the natural predictions of the discretized geometries and providing the microscopic interpretation of BH entropy \cite{Rovelli:1994ge,Ashtekar:1996eg,Ashtekar:1997fb,Yang:2016kia,Thiemann:1996at,Ma:2010fy,Ashtekar:1997yu,Song:2020arr}. The nonperturbative quantization procedure of LQG has been successfully applied to the metric $f(R)$ theories \cite{Zhang:2011vi,Zhang:2011qq}, scalar-tensor theories \cite{Zhang:2011gn,Zhang:2011vg}, higher-dimensional gravity \cite{Bodendorfer:2011nv}, and so on \cite{Zhang:2020smo}. Despite these achievements, the dynamics of full LQG is still an unsolved issue. To gain a certain level of understanding of the dynamics, the quantization ideas and technologies developed in LQG have also be applied to its symmetry-reduced models, such as the Friedmann-Robertson-Walker (FRW) models and the spherically symmetric black hole models, leading to loop quantum cosmology (LQC) and loop quantum black hole models \cite{Ashtekar:2003hd,Ashtekar:2005qt}. The most successful feature of LQC is that it can resolve the classical big bang singularity by a quantum bounce due to the quantum geometry effects. We refer to \cite{Ashtekar:2003hd,Bojowald:2005epg,Ashtekar:2006wn,Ashtekar:2011ni} for more complete reviews on LQC.

In full LQG, the gravitational Hamiltonian constraint is a combination of the so-called Euclidean term and the Lorentzian term. In the spatially flat, $k=0$ FRW model, the Lorentzian term and the Euclidean term are proportional to each other. Thus one often combines these two terms into one term proportional to the Euclidean term, and then quantizes the Euclidean term to obtain the well-defined gravitational Hamiltonian constraint operator \cite{Ashtekar:2003hd,Ashtekar:2006wn}. It turns out that in this quantization scheme the classical big-bang singularity is replaced by a symmetric quantum bounce for the $k=0$ FRW model with a massless scalar field in the framework of LQC \cite{Ashtekar:2006wn}. Note that in full LQG, the Lorentzian term is quantized independently by employing the Thiemann's trick \cite{Thiemann:1996aw}. Thus, to mimic the full LQG quantization procedure in the $k=0$ model of LQC, the Euclidean term and the Lorentzian term were treated independently \cite{Bojowald:2002gz,Henriques:2006qb,Yang:2009fp}. This alternative quantization scheme leads to an asymmetric quantum bounce, which relates the spatially flat FRW model with an asymptotic de Sitter universe, and thus an effective cosmological constant and an effective Newtonian constant can be obtained \cite{Assanioussi:2018hee,Li:2018opr,Zhang:2021zfp}.

As in the $k=0$ model, the quantization technologies for the gravitational Hamiltonian constraint developed in LQG have been extended to the $k=-1,+1$ models \cite{Vandersloot:2006ws,Szulc:2007uk,Szulc:2006ep,Ashtekar:2006es,Mielczarek:2009kh,Corichi:2011pg,Corichi:2013usa,Dupuy:2016upu}. Compared to the $k=0$ model where the spin connection vanishes and hence the Ashtekar connection equals to the extrinsic curvature multiplied by the Immirzi parameter, the Lorentzian term is not proportional to the total Euclidean term, but is proportional to the part of the Euclidean term involving the extrinsic curvature due to the nonvanishing spin connection for both the $k=-1$ model and the $k=+1$ model. Hence in the literatures one often absorbs the Lorentzian term into a part of the Euclidean term, and then quantize the two parts of the Euclidean term, respectively. It turns out that, as the $k=0$ model with the similar treatment, the resulting $k=-1$ LQC model also predicts a vacuum repulsion in the high curvature regime that would lead to a symmetric bounce \cite{Vandersloot:2006ws}. Moreover, the $k=-1$ model of LQC also possesses some new features that never appears in the $k=0$ model, for example, due to a vacuum repulsion in the high curvature regime, the scale factor has the minimum value as $a_{\rm min}=\gamma\sqrt{\Delta}$ \cite{Vandersloot:2006ws}. It is natural to ask whether the treatment of the Lorentzian term independently, mimicking the treatment in the full theory, can be directly carried to the $k=-1$ model, and whether an asymmetric bounce can still be held for the $k=-1$ model. This is the main motivation of the present paper. In this paper, we consider an alternative quantization of the gravitational Hamiltonian constraint in the $k=-1$ model by treating the Lorentzian term independently.

This paper is organized as follows. The canonical formulation of the $k=-1$ model is briefly recalled in Sec. \ref{section2}. Then we propose an alternative gravitational Hamiltonian constraint operator by treating the Lorentzian term independently, and provide a new quantum dynamics for the $k=-1$ model in Sec. \ref{section3}. The effective theory of the new quantum dynamics and its asymptotic behavior are studied in Sec. \ref{section4}. Summary is included in the last section.

\section{canonical formulation of the $k=-1$ model}\label{section2}

According to the cosmological principle, the line elements of the homogenous isotropic cosmological models take as
\begin{align}
{\rm d}s^2=-{\rm d}t^2+a^2(t)\left[\frac{1}{1-kr^2}{\rm d}r^2+r^2\left({\rm d}\theta^2+\sin^2\theta {\rm d}\phi^2\right)\right],
\end{align}
where $a(t)$ is the scale factor, and $k=-1,0,1$ for the open, flat, and closed FRW models, respectively.

In what follows, we present the canonical formulation of the $k=-1$ model following Ref. \cite{Vandersloot:2006ws}. For the spatially noncompact $k=0,-1$ models with topology homeomorphic to $\mathbb{R}^3$, one introduces an ``elemental cell" ${\cal V}$ on the homogeneous spatial manifold $\mathbb{R}^3$ and restrict all integrals to this elemental cell. Then one chooses a fiducial metric ${}^o\!q_{ab}={}^o\!\omega^i_a{}^o\!\omega^j_b\delta_{ij}$ on $\mathbb{R}^3$ with ${}^o\!\omega^i_a$ being the left- and right-invariant fiducial one-forms in the $k=0$ model, and only the left-invariant fiducial one-forms in the $k=-1$ model. Here $a,b,\cdots$ denote the spatial indices while $i,j,\cdots =1,2,3$. Denote by $V_o$ the volume of ${\cal V}$ measured by the fiducial metric ${}^o\!q_{ab}$. The left-invariant one-forms ${}^o\!\omega^i_a$ satisfy the Maurer-Cartan equation
\begin{align}
{\rm d}\ {}^o\!\omega^i+\frac{1}{2}{C^i}_{jk}\ {}^o\!\omega^j\wedge {}^o\!\omega^k=0,
\end{align}
where for the $k=-1$ model the structure constants read
\begin{align}
{C^i}_{jk}=\delta^i_j\delta_{k1}-\delta^i_k\delta_{j1},
\end{align}
while for the $k=0$ model they take zero. The corresponding left-invariant vector fields ${}^o\!e^a_i$ are dual to ${}^o\!\omega^i_a$, satisfying ${}^o\!e^a_i{}^o\!\omega^j_a=\delta^j_i$ and ${}^o\!e^a_i{}^o\!\omega^i_b=\delta^a_b$. The commutators between the left-invariant vector fields read
\begin{align}
 [{}^o\!e_i,{}^o\!e_j]={C^k}_{ij}{}^o\!e_k.
\end{align}

Classically, the dynamical variables of LQC are obtained by symmetrically reducing those of full LQG. In the full theory, the dynamical variables consist of the $su(2)$-valued connection $A_a^i$ and the densitized triad $\tilde{E}^b_j$ with the nontrivial Poisson bracket
\begin{align}
\{A^i_a(x),\tilde{E}_j^b(y)\}&=\kappa\gamma\delta^b_a\delta^i_j\delta(x,y),
\end{align}
where $\kappa=8\pi G$ with $G$ being the Newtonian constant, and $\gamma$ is the Immirzi parameter \cite{Barbero:1994ap,Immirzi:1996dr}. The connection $A^i_a$ is related to the spin connection $\Gamma^i_a$ and the extrinsic curvature $K^i_a$ by $A^i_a=\Gamma^i_a+\gamma K^i_a$. It turns out that the symmetry-reduced extrinsic curvature $K^i_a$ is diagonal in the basis of left-invariant one-forms for the $k=0,-1$ models. While, unlike the $k=0$ model where $\Gamma^i_a$ vanishes, the symmetry-reduced spin connection $\Gamma^i_a$ in the $k=-1$ model takes the form \cite{Vandersloot:2006ws}
\begin{align}\label{eq:gamma-express}
\Gamma^i_a=-\epsilon^{1ij}\ {}^o\!\omega^j_a,
\end{align}
and thus it is nondiagonal. Hence the symmetry-reduced connection and densitized triad for the $k=-1$ model read \cite{Vandersloot:2006ws}
\begin{align}\label{eq:A-E-forms-A}
 A_a^i&=-\epsilon^{1ij}\ {}^o\!\omega^j_a+cV_o^{-\frac13}\ {}^o\!\omega^i_a\equiv A^i_j V_o^{-\frac13}\ {}^o\!\omega^j_a,\\
\tilde{E}^a_i&=pV_o^{-\frac23}\sqrt{\det({}^o\!q)}\ {}^o\!e^a_i,\label{eq:A-E-forms-E}
\end{align}
where
\begin{align}
 A^i_j=
 \begin{pmatrix}
c & 0 & 0\\
0 & c & -V_o^{\frac{1}{3}}\\
0 & V_o^{\frac{1}{3}} & c
\end{pmatrix},
\end{align}
the variables $c$ and $p$ are only functions of $t$, and $\det({}^o\!q)$ denotes the determinant of ${}^o\!q_{ab}$. Hence the gravitational phase space of the $k=-1$ model consists of conjugate pairs $(c,p)$. The nontrivial Poisson bracket reads
\begin{align}\label{poissonb}
\{c,p\}&=\frac{\kappa}{3}\gamma.
\end{align}
Note that the variables $c$ and $p$ are related to the scale factor $a$ by $|p|=a^2V_o^{\frac 23}$ and $c=\gamma\dot{a} V_o^{\frac 13}$. The physical volume $V$ of the elemental cell ${\cal V}$ measured by the spatial (physical) metric $q_{ab}=|p|V_o^{-\frac 23}{}^o\!q_{ab}$ is related to $p$ via $V={|p|}^{3/2}$. In the improved scheme, it is convenient to choose the following variables to simplify the dynamics \cite{Ding:2008tq}
\begin{align}
b&:=\frac{\bar\mu c}{2}, \qquad v:=\frac{{\rm sgn}(p)|p|^{3/2}}{2\pi\gamma\ell^2_{\rm p}\sqrt\Delta},
\end{align}
where $\ell_{\rm p}\equiv\sqrt{G\hbar}$ denotes the Planck length, ${\rm sgn}(p)$ is the signature of $p$, $\Delta\equiv 4\sqrt{3}\,\pi\gamma\ell^2_{\rm p}$ is the minimum nonzero eigenvalue of the area operator in full LQG  \cite{Ashtekar:2008zu}, and  $\bar\mu\equiv \sqrt{\Delta/|p|}$. The Poisson bracket between $b$ and $v$ is given by
\begin{align}\label{eq:poisson-bracke}
\{b,v\}=\frac{1}{\hbar}.
\end{align}
As in the $k=0$ model, the Gauss and diffeomorphism constraints of the gravitational part are automatically satisfied for the symmetry-reduced variables in Eqs. \eqref{eq:A-E-forms-A} and \eqref{eq:A-E-forms-E} in the $k=-1$ model, and thus the classical dynamics is encoded in the Hamiltonian constraint. The gravitational Hamiltonian constraint of the $k=-1$ model reads
\begin{widetext}
\begin{align}\label{classical-Hamiltonian-minus}
{\cal H}^{k=-1}_{\rm grav}&:=\int_{\cal V}{\rm d}^3x\, \frac{\tilde{E}^a_i\tilde{E}^b_j}{2\kappa\sqrt{\textrm{det}(q)}}\left[{\epsilon^{ij}}_k{}^{\sst(A)}\!F^{k}_{ab}-2(1+\gamma^2)K^i_{[a}K^j_{b]}\right]\notag\\
&=\int_{\cal V}{\rm d}^3x\, \frac{\tilde{E}^a_i\tilde{E}^b_j}{2\kappa\sqrt{\textrm{det}(q)}}\left[{\epsilon^{ij}}_k{}^{\sst(\gamma K)}\!F^{k}_{ab}-2(1+\gamma^2)K^i_{[a}K^j_{b]}\right]+\int_{\cal V}{\rm d}^3x\, \frac{\tilde{E}^a_i\tilde{E}^b_j}{2\kappa\sqrt{\textrm{det}(q)}}{\epsilon^{ij}}_k{}^{(\Gamma)}\!F^{k}_{ab}\notag\\
&\equiv\tilde{\cal H}^{{\rm E},k=0}_{\rm grav}-2(1+\gamma^2)\tilde{\cal H}^{{\rm L},k=0}_{\rm grav}+{\cal H}_{\rm grav}^{\Gamma,k=-1},
\end{align}
\end{widetext}
where $\det(q)$ denotes the determinant of $q_{ab}$, and
\begin{align}
 {}^{\sst(x)}\!F^{k}_{ab}:=2\partial_{[a}x^k_{b]}+{\epsilon^k}_{lm}x^l_ax^m_b.
\end{align}
Hence, in the $k=-1$ model, the Euclidean term consists of $\tilde{\cal H}^{{\rm E},k=0}_{\rm grav}$ and ${\cal H}_{\rm grav}^{\Gamma,k=-1}$, while the Lorentzian term is $\tilde{\cal H}^{{\rm L},k=0}_{\rm grav}$. Seen from the above formulation, the two terms $\tilde{\cal H}^{{\rm E},k=0}_{\rm grav}$ and $\tilde{\cal H}^{{\rm L},k=0}_{\rm grav}$ have the same formulations as the Euclidean term ${\cal H}^{{\rm E},k=0}_{\rm grav}$ and the Lorentzian term ${\cal H}^{{\rm E},k=0}_{\rm grav}$ in the $k=0$ model, respectively. Thus the gravitational Hamiltonian constraint of the $k=-1$ model differs from that of the $k=0$ model by the third term ${\cal H}_{\rm grav}^{\Gamma,k=-1}$ due to the nonvanishing  $\Gamma^i_a$ in the $k=-1$ model. A straightforward calculation shows that the three terms in Eq. \eqref{classical-Hamiltonian-minus} can be expressed by the variables $(b,v)$ as
\begin{align}
\tilde{\cal H}_{\rm grav}^{{\rm E},k=0}&=\frac{3\gamma\hbar}{\sqrt{\Delta}}\,b^2|v|,\label{eq:Euclidean-term}\\
\tilde{\cal H}_{\rm grav}^{{\rm L},k=0}&=\frac{3\hbar}{2\gamma\sqrt{\Delta}}\,b^2|v|,\label{eq:Lorentz-term}\\
{\cal H}^{\Gamma,k=-1}_{\rm grav}&=\frac{3\left(\gamma\sqrt{\Delta}\,\hbar\right)^{\frac13}V_o^{\frac23}}{4\left(2\pi G\right)^{\frac23}}|v|^{\frac13}.
\end{align}
Hence the gravitational Hamiltonian constraint \eqref{classical-Hamiltonian-minus} of the $k=-1$ model reduces to
\begin{align}
 {\cal H}^{k=-1}_{\rm grav}&=-\frac{3\hbar|v|}{\gamma\sqrt{\Delta}}\left[b^2-V_o^{\frac23}\left(\frac{\gamma^2\Delta}{16\pi G\hbar|v|}\right)^{\frac23}\right]\notag\\
 &\equiv-\frac{3\hbar|v|}{\gamma\sqrt{\Delta}}g(b,v).
\end{align}

At the classical level, we assume that the universe is filled by a massless scalar field $\phi$. The Hamiltonian of the scalar field $\phi$ is given by
\begin{align}\label{eqn:matter-hamilton}
{\cal H }_\phi=\frac{p^2_\phi}{2V}=\frac{p_\phi^2}{4\pi\gamma\ell_{\rm P}^2\sqrt{\Delta}\,|v|},
\end{align}
where $p_\phi$ denotes the conjugate momentum of $\phi$.  The Poisson bracket between $\phi$ and $p_\phi$ is $\{\phi,p_\phi\}=1$. Hence the total Hamiltonian constraint of gravity coupled to a massless scalar field reads
\begin{align}\label{eq:class-tot-Ham}
 {\cal H}_{\rm tot}^{k=-1}&={\cal H}^{k=-1}_{\rm grav}+{\cal H}_{\phi}\notag\\
 &=-\frac{3\hbar|v|}{\gamma\sqrt{\Delta}}g(b,v)+\frac{p_\phi^2}{4\pi\gamma\ell_{\rm P}^2\sqrt{\Delta}\,|v|}.
\end{align}
By the total Hamiltonian constraint equation
\begin{align}\label{eq:constraint-eq-class}
 {\cal H}_{\rm tot}^{k=-1}=0,
\end{align}
the classical Friedmann equation can be obtained as
\begin{align}\label{eq:classical-Hubble-expression-k-negative}
 H^2_{k=-1}&=\left(\frac{\dot{v}}{3v}\right)^2=\left(\frac{\{v,{\cal H}_{\rm tot}^{k=-1}\}}{3v}\right)^2\notag\\
 &=\frac{8\pi G}{3}\rho_\phi+\frac{V_o^{2/3}}{V^{2/3}}\notag\\
 &=\frac{8\pi G}{3}\rho_\phi+\frac{1}{a^2},
\end{align}
where $\cdot{}$ denotes a derivative with respect to the time determined by ${\cal H}_{\rm tot}^{k=-1}$, and $\rho_\phi=\frac{p_\phi^2}{2V^2}$ is the energy density of the scalar field $\phi$.

\section{Loop quantization of the $k=-1$ model} \label{section3}

To pass the classical theory of $k=-1$ model to its quantum theory, one needs to construct the kinematical Hilbert space. In the $k=0$ model, the vanishing $\Gamma^i_a$ enables us to identify $A^i_a$ with $\gamma K^i_a$, leading to the identification of the holonomies of the connection and those of the extrinsic curvature (mutiplied by $\gamma$). The resulting holonomies of the connection $A^i_a$, equal to $\gamma K^i_a$ in the $k=0$ model, along edges generated by the left- and right-invariant vector fields ${}^o\!e^a_i$ with physical length $\lambda V^{1/3}$ take the form $h^{(\lambda)}_i=\cos \left(\frac{\lambda c}{2}\right)\mathbb{I}+2\tau_i\sin \left(\frac{\lambda c}{2}\right)$, where $\tau_i:=-\frac{{\rm i}}{2}\sigma_i$ with $\sigma_i$ being the Pauli matrices. Hence the related algebra is that of the almost periodic functions, and thus the kinematical Hilbert space for the gravitational part can be defined as ${\mathscr H}^{{\rm gr},k=0}_{\rm kin}=L^2(\mathbb{R}_{\mathrm{Bohr}},{\rm d}\mu_{\mathrm{Bohr}})$, where ${\mathbb R}_{\rm Bohr}$ and ${\rm d}\mu_{\rm Bohr}$ are respectively the Bohr compactification of the real line ${\mathbb R}$ and the Haar measure on it \cite{Ashtekar:2003hd}. However, in the $k=-1$ case, the spin connection $\Gamma^i_a$ takes the nonvanishing expression \eqref{eq:gamma-express}, resulting in a difference between the holonomy of the connection and the one of the extrinsic curvature. Moreover, due to the nondiagonal form \eqref{eq:A-E-forms-A} of the connection, the holonomies of the connection take complicated forms in the $k=-1$ model, leading to the algebra generated is no longer that of the almost periodic function \cite{Vandersloot:2006ws}. Instead, one often considers the holonomies of the extrinsic curvature $\gamma K^i_a$ in the $k=-1$ model, which take the same forms as those in the $k=0$ model. More precisely, considering an edge $e_i$ starting from the basepoint of the elemental cell ${\cal V}$, with tangent vector parallel to the vector ${}^o\!e^a_i$ and taking length $\lambda$, following Refs. \cite{Vandersloot:2006ws,Szulc:2007uk}, we define the ``holonomy" of $\gamma K^i_a=cV_o^{-\frac13}\ {}^o\!\omega^i_a$ as
\begin{align}\label{eq:lambda-holonomy}
h^{(\lambda)}_i&:={\cal P}\exp\int_{e_i} {\rm d}t\,\gamma K^j_a\tau_j{}^o\!e^a_i\notag\\
&=e^{\lambda c\tau_i}\notag\\
&=\cos \left(\frac{\lambda c}{2}\right)\mathbb{I}+2\sin\left(\frac{\lambda c}{2}\right)\tau_i.
\end{align}
Here ${\cal P}$ denotes the path ordering which orders the smallest path parameter to the left \cite{Thiemann:2007pyv}, and it takes the trivial action in our model as in the works \cite{Thiemann:2007pyv,Ashtekar:2003hd,Vandersloot:2006ws,Szulc:2007uk}. Clearly, these holonomies \eqref{eq:lambda-holonomy} generate the algebra of almost periodic functions, and thus result in the kinematical Hilbert space for the $k=-1$ model being ${\mathscr H}^{\rm gr}_{\rm kin}\equiv{\mathscr H}^{{\rm gr},k=-1}_{\rm kin}={\mathscr H}^{{\rm gr},k=0}_{\rm kin}$ \cite{Vandersloot:2006ws}. As in the $k=0$ model, we will employ the $\bar\mu$-scheme to define the Hamiltonian operator. This requires us to consider the holonomis along the edges taking physical length $\sqrt{\Delta}$ \cite{Thiemann:2007pyv,Vandersloot:2006ws,Szulc:2007uk}, which are given by
\begin{align}\label{eqn:i-holonomy}
 h^{(\bar\mu)}_i&:={\cal P}\exp\left(\int_0^{\bar{\mu}V_o^{1/3}} {\rm d}t\,\gamma K^j_a\tau_j{}^o\!e^a_i\right)\notag\\
 &=e^{\bar{\mu}c\tau_i}\notag\\
 &=\cos\left(\frac{\bar{\mu}c}{2}\right)\,\mathbb{I}+2\sin\left(\frac{\bar{\mu}c}{2}\right)\tau_i\notag\\
&=\cos(b)\,\mathbb{I}+2\sin(b)\,\tau_i\,,
\end{align}
and their inverse take the forms:
\begin{align}
 {h_i^{(\bar{\mu})}}^{-1}=\cos(b)\,\mathbb{I}-2\sin(b)\,\tau_i.
\end{align}
In the $v$-representation for both the $k=0$ model and the $k=-1$ model, the two elementary operators, $\widehat{e^{{\rm i}b}}$ and $\hat{v}$, act on the basis $|v\rangle$ of ${\mathscr H}^{\rm gr}_{\rm kin}$ as
\begin{align}
\widehat{e^{{\rm i}b}}\,|v\rangle&=|v+1\rangle\,,\qquad\hat{v}\,|v\rangle=v\,|v\rangle\,.
\end{align}
Thus one can easily write down the action of the operators
\begin{align}
 \widehat{h^{(\bar\mu)}_i}=\widehat{\cos(b)}\,\mathbb{I}+2\,\widehat{\sin(b)}\,\tau_i
\end{align}
corresponding to the holonomies $h^{(\bar{\mu})}_i$ of the extrinsic curvature $\gamma K^i_a$ on $|v\rangle$ in terms of $\widehat{e^{{\rm i}b}}$. For the scalar field, it is convenient to choose the Schr\"odinger representation \cite{Ashtekar:2011ni}. Thus the kinematical Hilbert space for the scalar field part can be chosen as ${\mathscr H}_{{\rm kin}}^{\rm sc}:=L^2({\mathbb R},d\phi)$. Hence the total kinematical Hilbert space of the $k=-1$ model with a scalar field is ${\mathscr H}_{\rm kin}^{\rm tot}={\mathscr H}^{\rm gr}_{\rm kin}\otimes{\mathscr H}^{\rm sc}_{\rm kin}$.

We now consider an alternative regularization of the gravitational Hamiltonian constraint of the $k=-1$ model in Eq. \eqref{classical-Hamiltonian-minus}, such that it is closer to that in the $k=0$ model as well as to that in full LQG. As mentioned previously, the two terms $\tilde{\cal H}^{{\rm E},k=0}_{\rm grav}$ and $\tilde{\cal H}^{{\rm L},k=0}_{\rm grav}$ in Eq. \eqref{classical-Hamiltonian-minus} have the same forms as the Euclidean and Lorentzian terms in the $k=0$ model, respectively. Hence it is natural to expect that the two terms in the $k=-1$ model can be regularized as the corresponding forms in the $k=0$ model. To realize explicitly this idea, some subtle issues should be clarified. Firstly, we consider the first term
\begin{align}\label{eq:H-E-reduce}
 \tilde{\cal H}^{{\rm E},k=0}_{\rm grav}&=\int_{\cal V}{\rm d}^3x\, \frac{\tilde{E}^a_i\tilde{E}^b_j}{2\kappa\sqrt{\textrm{det}(q)}}{\epsilon^{ij}}_k{}^{\sst(\gamma K)}\!F^{k}_{ab}\notag\\
 &=\int_{\cal V}{\rm d}^3x\, \frac{\tilde{E}^a_i\tilde{E}^b_j}{2\kappa\sqrt{\textrm{det}(q)}}{\epsilon^{ij}}_k\left[2\partial_{[a}\gamma K^k_{b]}\right.\notag\\
 &\hspace{4cm}\left.+{\epsilon^k}_{lm}\gamma K^l_a\gamma K^m_b\right]\notag\\
 &=\int_{\cal V}{\rm d}^3x\, \frac{\tilde{E}^a_i\tilde{E}^b_j{\epsilon^{ij}}_k}{2\kappa\sqrt{\textrm{det}(q)}}{\epsilon^k}_{lm}\gamma K^l_a\gamma K^m_b,
\end{align}
where in the third step we used the fact that the first term in the integral of the second line vanishes due to
\begin{align}
  {}^o\!e^a_i{}^o\!e^b_j{\epsilon^{ij}}_k2\partial_{[a}\gamma K^k_{b]}&=cV_o^{-\frac13}{}^o\!e^a_i{}^o\!e^b_j{\epsilon^{ij}}_k2\partial_{[a}{}^o\!\omega^k_{b]}\notag\\
  &=-cV_o^{-\frac13}{}^o\!e^a_i{}^o\!e^b_j{\epsilon^{ij}}_k{C^k}_{lm}{}^o\!\omega^l_a{}^o\!\omega^m_b\notag\\
  &=-cV_o^{-\frac13}{\epsilon^{ij}}_k{C^k}_{ij}\notag\\
  &=0.
\end{align}
To regularize $\tilde{\cal H}^{{\rm E},k=0}_{\rm grav}$ in Eq. \eqref{eq:H-E-reduce}, one needs to use the Thiemann's trick
\begin{align}\label{eq:Thiemann-trick}
 \frac{\tilde{E}^a_i\tilde{E}^b_j}{\sqrt{\textrm{det}(q)}}{\epsilon^{ij}}_k&=\frac{2}{\kappa\gamma}\tilde{\epsilon}^{abc}\{A^k_c,V\}=\frac{2}{\kappa\gamma}\tilde{\epsilon}^{abc}\{\gamma K^k_c,V\},
\end{align}
where $\tilde{\epsilon}^{abc}$ is the Levi-Civita density, and then to express the curvature ${}^{\sst(\gamma K)}\!F^{k}_{ab}$ of $\gamma K^i_a$ in terms of holonomies of $\gamma K^i_a$. In the $k=0$ case, since the left- and right-invariant vector fields ${}^o\!e^a_i$ commute to each other, the integral curvatures of ${}^o\!e^a_i$ and ${}^o\!e^b_j$ can form closed loops $\Box_{ij}$, around which the curvature ${}^{\sst(\gamma K)}\!F^{k}_{ab}$ can be recast as the holonomies $h^{(\bar{\mu})}_{\Box_{ij}}$ of $\gamma K^i_a$. Compared to the $k=0$ case, due to the noncommutativity of the left-invariant vector fields ${}^o\!e^a_i$ in the $k=-1$ case, the integral curves of ${}^o\!e^a_i$ and ${}^o\!e^a_j$ can not provide closed loops. In \cite{Vandersloot:2006ws}, holonomies of $\gamma K^i_a$ based on the open curves generated by ${}^o\!e^a_i$ and ${}^o\!e^a_j$ were adopted to regularize the curvature ${}^{\sst(\gamma K)}\!F^{k}_{ab}$. In \cite{Szulc:2007uk}, the author proposed closed loops $\Box_{ij}$ generated by the integral curves of the left-invariant vector fields ${}^o\!e^a_i$ and the right-invariant vector fields ${}^o\!\eta^b_j$ commuting with ${}^o\!e^a_i$. In the present paper, we will consider open holonnomies to represent extrinsic curvature following Refs. \cite{Yang:2009fp,Ashtekar:2009um,Corichi:2011pg,Dupuy:2016upu}. Inputting Eq. \eqref{eq:Thiemann-trick} into Eq. \eqref{eq:H-E-reduce}, one obtains
\begin{align}\label{eq:H-E-reduce-one}
 \tilde{\cal H}^{{\rm E},k=0}_{\rm grav}&=\frac{1}{\kappa^2\gamma}\int_{\cal V}{\rm d}^3x\,\tilde{\epsilon}^{abc}{\epsilon^k}_{lm}\gamma K^l_a\gamma K^m_b\{\gamma K^k_c,V\}\notag\\
 &=-\frac{4}{\kappa^2\gamma}\int_{\cal V}{\rm d}^3x\,\tilde{\epsilon}^{abc}{\rm Tr}\left(\gamma K_a\gamma K_b\{\gamma K_c,V\}\right),
\end{align}
where the identity ${\rm Tr}(\tau_i\tau_j\tau_k)=-\frac{1}{4}\epsilon_{ijk}$ was used. In cosmology the known identities \cite{Yang:2009fp,Ashtekar:2009um,Corichi:2011pg,Dupuy:2016upu} take the forms:
\begin{align}
 \gamma K_a&=\frac{h^{(2\bar{\mu})}_i-{h_i^{(2\bar{\mu})}}^{-1}}{4\bar{\mu}V_o^{1/3}}{}^o\!\omega^i_a,\label{eq:id-1}\\
 \{\gamma K_c,V\}&=-\frac{1}{\bar{\mu}V_o^{1/3}}\sum_kh^{(\bar{\mu})}_k\left\{{h_k^{(\bar{\mu})}}^{-1},V\right\}{}^o\!\omega^k_c\label{eq:id-2},
\end{align}
where $h_i^{(\bar{\mu})}$ (or $h^{(2\bar{\mu})}_i$) is defined by \eqref{eqn:i-holonomy}.
It should be noticed that Eq. \eqref{eq:id-1}, which is precisely valid in the limit $\bar{\mu}\rightarrow0$, should be understood as a regularized expression in the ${\bar{\mu}}$-scheme with $\bar{\mu}=\sqrt{\Delta/|p|}$. Substituting Eqs. \eqref{eq:id-1} and \eqref{eq:id-2} into Eq. \eqref{eq:H-E-reduce-one} and assuming for simplicity that the holonomies of $k=-1$ can be approximated with the holonomies of $k=0$, we arrive at
\begin{widetext}
\begin{align}\label{eq:regularized-Euclidean-k-1}
  \tilde{\cal H}^{{\rm E},k=0,{\rm reg}}_{\rm grav}&=\frac{{\rm sgn}(p)}{4\kappa^2\gamma\bar{\mu}^3}\sum_{i,j,k}\epsilon^{ijk}{\rm Tr}\left[\left(h^{(2\bar{\mu})}_i-{h_i^{(2\bar{\mu})}}^{-1}\right)\left(h^{(2\bar{\mu})}_j-{h_j^{(2\bar{\mu})}}^{-1}\right)h^{(\bar{\mu})}_k\left\{{h_k^{(\bar{\mu})}}^{-1},V\right\}\right]\notag\\
  &=\frac{\hbar^2\gamma}{4\sqrt{\Delta}}\sin(2b)\left[v\sum_k{\rm Tr}\left(\tau_kh^{(\bar{\mu})}_k\left\{{h_k^{(\bar{\mu})}}^{-1},|v|\right\}\right)\right]\sin(2b),
\end{align}
\end{widetext}
where the identity $\tau_i\tau_j=\frac12\epsilon_{ijm}\tau^m-\frac14\delta_{ij}$ was used. The resulting regularized expression $\tilde{\cal H}^{{\rm E},k=0,{\rm reg}}_{\rm grav}$ in Eq. \eqref{eq:regularized-Euclidean-k-1} is the same as the regularized Euclidean Hamiltonian constraint ${\cal H}^{{\rm E},k=0,{\rm reg}}_{\rm grav}$ in the $k=0$ model \cite{Ashtekar:2006wn}. We now consider the second term $\tilde{\cal H}^{{\rm L},k=0}_{\rm grav}$ in Eq. \eqref{classical-Hamiltonian-minus}. Classically, the term $\tilde{\cal H}^{{\rm L},k=0}_{\rm grav}$ is proportional to the term $\tilde{\cal H}^{{\rm E},k=0}_{\rm grav}$, and hence the term $\tilde{\cal H}^{{\rm L},k=0}_{\rm grav}$ does not need to be quantized independently. This approach to quantization of $\tilde{\cal H}^{{\rm L},k=0}_{\rm grav}$ in the $k=-1$ has been adopted in \cite{Vandersloot:2006ws,Szulc:2007uk}, similar to the $k=0$ case in \cite{Ashtekar:2006wn}. Alternatively, the Lorentzian term in the $k=0$ case can be regularized independently in \cite{Yang:2009fp}, mimicking the treatment of full LQG. It is natural to ask whether the treatment for the Lorentzian term in the $k=0$ case can be directly carried to that for $\tilde{\cal H}^{{\rm L},k=0}_{\rm grav}$ in the $k=-1$ model, and the resulting operator is the same as that in the $k=0$ model. The answer is in the affirmative. To this end, let us recall the key identities for regularizing the Lorentzian term of the gravitational Hamiltonian constraint in the full theory, study their symmetry-reduced forms in the $k=-1$ model, and then compare them with those in the $k=0$ model. The first classical identity reads
\begin{align}\label{eq:H-L-id-1}
K_a^i\tau_i&=\frac{1}{\kappa\gamma}\{A^i_a\tau_i,K\}=\frac{1}{\kappa\gamma}\{\Gamma^i_a\tau_i+\gamma K^i_a\tau_i,K\}\notag\\
&=\frac{1}{\kappa\gamma}\{\gamma K^i_a\tau_i,K\}\notag\\
&=-\frac{2}{3\kappa\gamma}\frac{1}{\bar{\mu}V_o^{1/3}}\sum_ih^{(\bar{\mu})}_i\left\{{h_i^{(\bar{\mu})}}^{-1},K\right\}{}^o\!\omega^i_a,
\end{align}
where $K:=\int {\rm d}^3x K_a^i\tilde{E}^a_i$, the former steps hold for the full theory and thus hold for its symmetry-reduced models, while the third step holds due to the fact that the spin connection $\Gamma^i_a$ in Eq. \eqref{eq:gamma-express} is proportional to ${}^o\!\omega^j_a$ up to a constant for the $k=-1$ case, and the vanishing connection for the $k=0$ model. In the last step, we have used the relation \cite{Yang:2009fp}
\begin{align}\label{eq:id-c-K}
 \{c\tau_i,K\}&=-\frac{2}{3\bar{\mu}}h^{(\bar{\mu})}_i\left\{{h_i^{(\bar{\mu})}}^{-1},K\right\}.
\end{align}
Here, it is worth noting that the above equation is satisfied only in the $\bar{\mu}$ scheme where $\bar{\mu}$ is a function of $p$, rather that a certain constant in the $\mu_o$ scheme. In the $\bar{\mu}$ scheme, $\bar{\mu}$ depending on $p$ does not commute with $K$, leading to a factor $2/3$ on the right-hand side of Eq. \eqref{eq:id-c-K}. The second identity is
\begin{align}\label{eq:H-L-id-2}
K&=\frac{1}{\gamma^2}\{{}^{(A)}{\cal H}^{\rm E},V\}=\frac{1}{\gamma^2}\{{}^{(\Gamma)}{\cal H}^{\rm E}+{}^{(\gamma K)}{\cal H}^{\rm E},V\}\notag\\
&=\frac{1}{\gamma^2}\{{}^{(\gamma K)}{\cal H}^{\rm E},V\}=\frac{1}{\gamma^2}\{\tilde{\cal H}^{{\rm E},k=0}_{\rm grav},V\}\notag\\
&=\frac{1}{\gamma^2}\{{\cal H}^{{\rm E},k=0}_{\rm grav},V\},
\end{align}
where ${}^{(x)}{\cal H}^{\rm E}:=\int{\rm d}^3x\, \frac{\tilde{E}^a_i\tilde{E}^b_j}{2\kappa\sqrt{\textrm{det}(q)}}{\epsilon^{ij}}_k{}^{\sst(x)}\!F^{k}_{ab}$. Hence, to regularize $K$, one just replaces $\tilde{\cal H}^{{\rm E},k=0}_{\rm grav}$ by its regularized version $\tilde{\cal H}^{{\rm E},k=0,{\rm reg}}_{\rm grav}$ in Eq. \eqref{eq:H-L-id-2}. It should also be noted here that both $\bar{\mu}$ and $V$ depend only on $p$. As a result, the Poisson bracket between $\tilde{\cal H}^{{\rm E},k=0,{\rm reg}}_{\rm grav}$ and $V$ has the same form for both the $\bar{\mu}$ and $\mu_o$ schemes. Thus the above two classical identities which play key roles in the regularization of the Lorentzian term hold in the $k=-1$ model, and take the same forms as those in the $k=0$ model. Therefore, to regularize $\tilde{\cal H}^{{\rm L},k=0}_{\rm grav}$ independently in the $k=-1$ model, one can follow directly the treatment of the Lorentzian term ${\cal H}^{{\rm L},k=0}_{\rm grav}$ in the $k=0$ model, mimicking the treatment of the full theory. To this end, we first re-express $\tilde{\cal H}^{{\rm L},k=0}_{\rm grav}$ in the form
\begin{align}\label{eq:H-L-reduce-one}
  \tilde{\cal H}^{{\rm L},k=0}_{\rm grav}&=\int_{\cal V}{\rm d}^3x\, \frac{\tilde{E}^a_i\tilde{E}^b_j}{2\kappa\sqrt{\textrm{det}(q)}}K^i_{[a}K^j_{b]}\notag\\
  &=\int_{\cal V}{\rm d}^3x\, \frac{\tilde{E}^a_i\tilde{E}^b_j}{2\kappa\sqrt{\textrm{det}(q)}}{\epsilon^{ij}}_k\frac12{\epsilon^k}_{lm}K^l_aK^m_b\notag\\
  &=-\frac{2}{\kappa^2\gamma}\int_{\cal V}{\rm d}^3x\,\tilde{\epsilon}^{abc}{\rm Tr}\left(K_aK_b\{\gamma K_c,V\}\right).
\end{align}
Combining Eq. \eqref{eq:H-L-reduce-one} with Eqs. \eqref{eq:H-L-id-1} and \eqref{eq:id-2} and replacing $\tilde{\cal H}^{{\rm E},k=0}_{\rm grav}$ by its regularized version $\tilde{\cal H}^{{\rm E},k=0,{\rm reg}}_{\rm grav}$,
one obtains the regularized expression of $\tilde{\cal H}^{{\rm L},k=0}_{\rm grav}$ as \cite{Yang:2009fp}
\begin{widetext}
\begin{align}\label{eq:regularized-Lorentzian-k-1}
 \tilde{\cal H}^{{\rm L},k=0,{\rm reg}}_{\rm grav}&=\frac{8\,{\rm sgn}(p)}{9\kappa^4\gamma^7\bar{\mu}^3}\sum_{i,j,k}\epsilon^{ijk}{\rm Tr}\left(h^{(\bar{\mu})}_i\left\{{h_i^{(\bar{\mu})}}^{-1},\left\{\tilde{\cal H}^{{\rm E},k=0,{\rm reg}}_{\rm grav},V\right\}\right\}h^{(\bar{\mu})}_j\left\{{h_j^{(\bar{\mu})}}^{-1},\left\{\tilde{\cal H}^{{\rm E},k=0,{\rm reg}}_{\rm grav},V\right\}\right\}h^{(\bar{\mu})}_k\left\{{h_k^{(\bar{\mu})}}^{-1},V\right\}\right)\notag\\
 &=\frac{8\,{\rm sgn}(p)}{9\kappa^4\gamma^7\bar{\mu}^3}\sum_{i,j,k}\epsilon^{ijk}{\rm Tr}\left(h^{(\bar{\mu})}_j\left\{{h_j^{(\bar{\mu})}}^{-1},\left\{\tilde{\cal H}^{{\rm E},k=0,{\rm reg}}_{\rm grav},V\right\}\right\}h^{(\bar{\mu})}_k\left\{{h_k^{(\bar{\mu})}}^{-1},V\right\}h^{(\bar{\mu})}_i\left\{{h_i^{(\bar{\mu})}}^{-1},\left\{\tilde{\cal H}^{{\rm E},k=0,{\rm reg}}_{\rm grav},V\right\}\right\}\right)\notag\\
 &=\frac{\hbar^4\sqrt{\Delta}}{288\gamma^3}\sum_{i,j,k}\epsilon^{ijk}{\rm Tr}\left(h^{(\bar{\mu})}_i\left\{{h_i^{(\bar{\mu})}}^{-1},\left\{\tilde{\cal H}^{{\rm E},k=0,{\rm reg}}_{\rm grav},|v|\right\}\right\}v\,h^{(\bar{\mu})}_j\left\{{h_j^{(\bar{\mu})}}^{-1},|v|\right\}h^{(\bar{\mu})}_k\left\{{h_k^{(\bar{\mu})}}^{-1},\left\{\tilde{\cal H}^{{\rm E},k=0,{\rm reg}}_{\rm grav},|v|\right\}\right\}\right).
\end{align}
Similarly, the last term in Eq. \eqref{classical-Hamiltonian-minus} can be regularized as \cite{Vandersloot:2006ws,Szulc:2007uk}
\begin{align}\label{eq:regularized-Gamma-k-1}
 {\cal H}^{\Gamma,k=-1,{\rm reg}}_{\rm grav}&=\frac{{\rm sgn}(p)V_o^{\frac23}}{2\kappa\pi G\gamma\bar{\mu}}\sum_k{\rm Tr}\left(\tau_kh^{(\bar{\mu})}_k\left\{{h_k^{(\bar{\mu})}}^{-1},V\right\}\right)\notag\\
 &=\frac{\hbar(\hbar\gamma\sqrt{\Delta})^{\frac13}V_o^{\frac23}}{4(2\pi G)^{\frac23}}{\rm sgn}(v)|v|^{\frac13}\sum_k{\rm Tr}\left(\tau_kh^{(\bar{\mu})}_k\left\{{h_k^{(\bar{\mu})}}^{-1},|v|\right\}\right).
\end{align}
\end{widetext}
Up to now, the three parts of the gravitational Hamiltonian constraint in Eq. \eqref{classical-Hamiltonian-minus} have been regularized as the expressions in Eqs. \eqref{eq:regularized-Euclidean-k-1}, \eqref{eq:regularized-Lorentzian-k-1} and \eqref{eq:regularized-Gamma-k-1}, which can be directly promoted to quantum operators by replacing the functions by the corresponding operators and replacing $\{\cdot,\cdot\}$ by $[\cdot,\cdot]/({\rm i}\hbar)$. Then $\tilde{\cal H}^{{\rm E},k=0,{\rm reg}}_{\rm grav}$ in \eqref{eq:regularized-Euclidean-k-1} can be quantized as
\begin{widetext}
\begin{align}\label{eq:k0-Eculdian-op}
 \hat{\cal H}^{{\rm E},k=0}_{\rm grav}&=-\frac{{\rm i}\hbar\gamma}{4\sqrt{\Delta}}\widehat{\sin(2b)}\left(\hat{v}\sum_k{\rm Tr}\left(\tau_k\widehat{h^{(\bar{\mu})}_k}\left[\widehat{{h_k^{(\bar{\mu})}}^{-1}},|\hat{v}|\right]\right)\right)\widehat{\sin(2b)}\notag\\
 &=\frac{{\rm i}\hbar\gamma}{2\sqrt{\Delta}}\widehat{\sin(2b)}\left(\hat{v}\,\hat{O}_{|\hat{v}|}\right)\widehat{\sin(2b)}\sum_k{\rm Tr}(\tau_k\tau_k)\notag\\
 &=-\frac{{\rm i}3\hbar\gamma}{4\sqrt{\Delta}}\,\widehat{\sin(2b)}\left(\hat{v} \,\hat{O}_{|\hat{v}|}\right)\widehat{\sin(2b)},
\end{align}
where in the second step we have used
\begin{align}\label{eq:hAhinv}
 \widehat{h^{(\bar{\mu})}_k}\left[\widehat{{h_k^{(\bar{\mu})}}^{-1}},\hat{B}\right]&=\hat{B}\,\mathbb{I}-\widehat{h^{(\bar{\mu})}_k}\hat{B}\,\widehat{{h_k^{(\bar{\mu})}}^{-1}}\notag\\
 &=\hat{B}\,\mathbb{I}-\left(\widehat{\cos(b)}\,\mathbb{I}+2\,\widehat{\sin(b)}\,\tau_k\right)\hat{B}\left(\widehat{\cos(b)}\,\mathbb{I}-2\,\widehat{\sin(b)}\,\tau_k\right)\notag\\
 &=\hat{B}\,\mathbb{I}-\left[\widehat{\cos(b)}\hat{B}\,\widehat{\cos(b)}\,\mathbb{I}-4\,\widehat{\sin(b)}\hat{B}\,\widehat{\sin(b)}\,\tau_k\tau_k+2\left(\widehat{\sin(b)}\hat{B}\,\widehat{\cos(b)}-\widehat{\cos(b)}\hat{B}\,\widehat{\sin(b)}\right)\tau_k\right]\notag\\
 &=\left(\hat{B}-\widehat{\sin(b)}\hat{B}\,\widehat{\sin(b)}-\widehat{\cos(b)}\hat{B}\,\widehat{\cos(b)}\right)\mathbb{I}-2\,\hat{O}_{\hat{B}}\,\tau_k,
\end{align}
here in the fourth step the identity $\tau_k\tau_k=-\frac14\,\mathbb{I}$ for $k=1,2,3$ was used, the operator $\hat{O}_{\hat{B}}$, depending on the operator $\hat{B}$, is defined by
\begin{align}\label{eq:hatO}
\hat{O}_{\hat{B}}&:=\widehat{\sin(b)}\hat{B}\,\widehat{\cos(b)}-\widehat{\cos(b)}\hat{B}\,\widehat{\sin(b)},
\end{align}
and ${\rm Tr}(\tau_k)=0$. In the last step in Eq. \eqref{eq:k0-Eculdian-op} we have used $\sum_k{\rm Tr}(\tau_k\tau_k)=-\frac32$. Similarly, the regularized expression \eqref{eq:regularized-Lorentzian-k-1} can be quantized as
\begin{align}\label{eq:k0-Lorentzian-op}
 \hat{\cal H}^{{\rm L},k=0}_{\rm grav}&=-\frac{{\rm i}\sqrt{\Delta}}{288\hbar\gamma^3}\sum_{i,j,k}\epsilon^{ijk}{\rm Tr}\left(\widehat{h^{(\bar{\mu})}_i}\left[\widehat{{h_i^{(\bar{\mu})}}^{-1}},\left[\hat{\cal H}^{{\rm E},k=0}_{\rm grav},|\hat{v}|\right]\right]\hat{v}\,\widehat{h^{(\bar{\mu})}_j}\left[\widehat{{h_j^{(\bar{\mu})}}^{-1}},|\hat{v}|\right]\widehat{h^{(\bar{\mu})}_k}\left[\widehat{{h_k^{(\bar{\mu})}}^{-1}},\left[\hat{\cal H}^{{\rm E},k=0}_{\rm grav},|\hat{v}|\right]\right]\right)\notag\\
 &=\frac{{\rm i}\sqrt{\Delta}}{36\hbar\gamma^3}\,\hat{O}_{[\hat{\cal H}^{{\rm E},k=0}_{\rm grav},|\hat{v}|]}\left(\hat{v} \,\hat{O}_{|\hat{v}|}\right)\hat{O}_{[\hat{\cal H}^{{\rm E},k=0}_{\rm grav},|\hat{v}|]}\sum_{i,j,k}\epsilon^{ijk}{\rm Tr}(\tau_i\tau_j\tau_k)\notag\\
 &=-\frac{{\rm i}\sqrt{\Delta}}{24\hbar\gamma^3}\,\hat{O}_{[\hat{\cal H}^{{\rm E},k=0}_{\rm grav},|\hat{v}|]}\left(\hat{v} \,\hat{O}_{|\hat{v}|}\right)\hat{O}_{[\hat{\cal H}^{{\rm E},k=0}_{\rm grav},|\hat{v}|]},
\end{align}
where in the second step we have used Eq. \eqref{eq:hAhinv} and used that the terms involving zero, one and two $\tau$ in the trace vanish due to $\sum_{ijk}
\epsilon^{ijk}=0$, $\sum_{ijk}
\epsilon^{ijk}{\rm Tr}(\tau_k)=0$, and $\sum_{ijk}
\epsilon^{ijk}{\rm Tr}(\tau_i\tau_j)=-\frac12\sum_{ijk}
\epsilon^{ijk}\delta_{ij}=0$. The operators $\hat{O}_{|\hat{v}|}$ and $\hat{O}_{[\hat{\cal H}^{{\rm E},k=0}_{\rm grav},|\hat{v}|]}$ are defined according to Eq. \eqref{eq:hatO} with $\hat{B}=|\hat{v}|$ and $\hat{B}=[\hat{\cal H}^{{\rm E},k=0}_{\rm grav},|\hat{v}|]$, respectively. Moreover, in the last step in Eq. \eqref{eq:k0-Lorentzian-op}, we have used $\sum_{i,j,k}\epsilon^{ijk}{\rm Tr}(\tau_i\tau_j\tau_k)=-\frac14\sum_{i,j,k}\epsilon^{ijk}\epsilon_{ijk}=-\frac{3}{2}$. Finally, the operator corresponding to ${\cal H}^{\Gamma,k=-1,{\rm reg}}_{\rm grav}$ in Eq. \eqref{eq:regularized-Gamma-k-1} reads if assuming again for simplicity that the holonomies of $k=-1$ can be approximated with the holonomies of $k=0$
\begin{align}\label{eq:k0-Gamma-op}
  \hat{\cal H}^{\Gamma,k=-1}_{\rm grav}&=\frac{(\hbar\gamma\sqrt{\Delta})^{\frac13}V_o^{\frac23}}{{\rm i}4(2\pi G)^{\frac23}}{\rm sgn}(\hat{v})|\hat{v}|^{\frac13}\sum_k{\rm Tr}\left(\tau_k\widehat{h^{(\bar{\mu})}_k}\left[\widehat{{h_k^{(\bar{\mu})}}^{-1}},|\hat{v}|\right]\right)\notag\\
  &=-\frac{(\hbar\gamma\sqrt{\Delta})^{\frac13}V_o^{\frac23}}{{\rm i}2(2\pi G)^{\frac23}}{\rm sgn}(\hat{v})|\hat{v}|^{\frac13}\hat{O}_{|\hat{v}|}\sum_k{\rm Tr}(\tau_k\tau_k)\notag\\
  &=\frac{3(\hbar\gamma\sqrt{\Delta})^{\frac13}V_o^{\frac23}}{{\rm i}4(2\pi G)^{\frac23}}{\rm sgn}(\hat{v})|\hat{v}|^{\frac13}\hat{O}_{|\hat{v}|}.
\end{align}
\end{widetext}
The actions of the operators $\hat{\cal H}^{{\rm E},k=0}_{\rm grav}$, $\hat{\cal H}^{{\rm L},k=0}_{\rm grav}$, and $\hat{\cal H}^{\Gamma,k=-1}_{\rm grav}$ on $|v\rangle$ read
\begin{align}\label{eq:quantum-Euclidean-Hamiltonian}
\hat{\cal H}^{{\rm E},k=0}_{\rm grav}|v\rangle&=E_+(v)|v+4\rangle+E_0(v)|v\rangle+E_-(v)|v-4\rangle,\\
\hat{\cal H}^{{\rm L},k=0}_{\rm grav}|v\rangle&=L_+(v)|v+8\rangle+L_0(v)|v\rangle+L_-(v)|v-8\rangle,\label{eq:quantum-Lorentzian-Hamiltonian}\\
\hat{\cal H}^{\Gamma,k=-1}_{\rm grav}|v\rangle&=\Gamma(v)|v\rangle,\label{eq:quantum-Gamma-Hamiltonian}
\end{align}
where
\begin{align}
E_+(v)&=\frac{3\gamma\hbar}{32\sqrt{\Delta}}(v+2)M_{1,3}(v),\\
E_-(v)&=E_+(v-4),\\
E_0(v)&=-E_+(v)-E_-(v),\\
L_+(v)&=-\frac{\sqrt{\Delta}}{192\gamma^3\hbar}(v+4)M_{-1,1}(v+4)\notag\\
&\hspace{2cm}\times G_-(v+4)G_+(v+4),\\
L_-(v)&=L_+(v-8),\\
L_0(v)&=-\frac{\sqrt{\Delta}}{192\gamma^3\hbar}\left\{(v+4) M_{-1,1}(v+4)[G_+(v)]^2\right.\notag\\
&\hspace{1.3cm}\left.+(v-4) M_{-1,1}(v-4)[G_-(v)]^2\right\},\\
\Gamma(v)&=\frac{3\left(\gamma\sqrt{\Delta}\,\hbar\right)^{\frac13}V_o^{\frac23}}{8\left(2\pi G\right)^{\frac23}}{\rm sgn}(v)|v|^{\frac13}M_{1,-1}(v).
\end{align}
Here
\begin{align}
M_{a,b}(v)&:=|v+a|-|v+b|,\\
G_\pm(v)&:=E_\pm(v-1) M_{0,\pm4}(v-1)\notag\\
&\hspace{1cm}-E_\pm(v+1) M_{0,\pm4}(v+1).
\end{align}
The function $M_{a,b}(v)$ satisfies
\begin{align}
  M_{0,|k|}(v)=-M_{0,-|k|}(v+|k|).
\end{align}
Hence the action of the $k=-1$ gravitational Hamiltonian constraint operator $\hat{\cal H}^{k=-1}_{\rm grav}$ on $|v\rangle$ reads
\begin{align}\label{eq:action-k-1model}
\hat{\cal H}^{k=-1}_{\rm grav}|v\rangle=&\tilde{L}_+(v)|v+8\rangle+E_+(v)|v+4\rangle\notag\\
&+\left[E_0(v)+\tilde{L}_0(v)+\Gamma(v)\right]|v\rangle\notag\\
&+E_-(v)|v-4\rangle+\tilde{L}_-(v)|v-8\rangle,
\end{align}
where $\tilde{L}_*:=-2(1+\gamma^2)L_*$, here $*=+,-,0$.

On the other hand, the Hamiltonian constraint ${\cal H}_\phi$ for the scalar field can be quantized as a well-defined operator $\hat{\cal H}_\phi$ in ${\mathscr H}^{\rm tot}_{\rm kin}$, and the action of $\hat{\cal H}_\phi$ on a quantum state $|\psi\rangle=\psi(v,\phi)|v,\phi\rangle$ with $|v,\phi\rangle\equiv|v\rangle\otimes|\phi\rangle\in{\mathscr H}^{\rm gr}_{\rm kin}\otimes{\mathscr H}^{\rm sc}_{\rm kin}$ is given by \cite{Ashtekar:2006wn}
\begin{align}\label{quantum-matter-Hamiltonian}
 \hat{\cal H}_\phi\cdot\psi(v,\phi)=-\frac{\hbar^2}{4\pi\gamma\sqrt{\Delta}\ell_{\mathrm p}^2}C(v)\partial^2_\phi\psi(v,\phi),
\end{align}
where
\begin{align}
C(v)\equiv \left(\frac{3}{2}\right)^3|v|
\left|{|v+1|^{{1}/{3}}}-{|v-1|^{{1}/{3}}}\right|^3.
\end{align}

Combining equations above, one can write down the resulting quantum Hamiltonian constraint equation corresponding to its classical one \eqref{eq:constraint-eq-class} as
\begin{align}\label{eq:quantum-equation}
 \hat{\cal H}_{\rm tot}^{k=-1}\cdot\psi(v,\phi)=\left(\hat{\cal H}^{k=-1}_{{\rm grav}}+\hat{\cal H}_\phi\right)\cdot\psi(v,\phi)=0,
\end{align}
which describes the quantum evolution of the coupled system with the scalar field $\phi$ as an emergent time.

\section{Effective theory and its asymptotic behavior of the alternative $k=-1$ LQC}\label{section4}

By constructing certain coherent states peaked at points of the classical phase space and computing the expectation value of the Hamiltonian constraint operator under the coherent states, one can obtain the corresponding effective Hamiltonian constraint. To this end, we first note that the symmetry-reduced phase space of the $k=-1$ model coincides with that of the $k=0$ model. Hence certain coherent states constructed for the $k=0$ model can be directly carried to the $k=-1$ model. A Gaussian coherent state peaked at a point $(b_o,v_o,\phi_o,p_\phi)$ in the classical phase space with spreads $\epsilon$ and $\sigma$ in the gravitational sector and scalar field sector takes the form {\cite{Ding:2008tq}
\begin{align}\label{eqn:dual-semiclassical-state}
\left(\Psi_{(b_o,v_o,\phi_o,p_\phi)}\right|&:=\int {\rm d}\phi\sum_{v\in\mathbb{R}}e^{-\frac{\epsilon^2}{2}(v-v_o)^2} e^{{\rm i}b_o(v-v_o)}\notag\\
&\hspace{1cm}\times  e^{-\frac{\sigma^2}{2}(\phi-\phi_o)^2}e^{\frac{\rm i}{\hbar}p_\phi(\phi-\phi_o)}(v|\otimes (\phi|,
\end{align}
and its shadow on the regular lattice with spacing one reads
\begin{align}\label{eqn:shadow-grav}
\left|\Psi\right\rangle&:=\int{\rm d}\phi\sum_{n\in\mathbb{Z}}e^{-\frac{\epsilon^2}{2}(n-v_o)^2}e^{-{\rm i}b_o(n-v_o)}\notag\\
&\hspace{1cm}\times e^{-\frac{\sigma^2}{2}(\phi-\phi_o)^2} e^{-\frac{\rm i}{\hbar}p_\phi(\phi-\phi_o)}\,|n\rangle\otimes |\phi\rangle\notag\\
&\equiv \left|\Psi_{\rm grav}\right\rangle\otimes \left|\Psi_{\phi}\right\rangle.
\end{align}
To make the state be sharply peaked in the classical phase space of the universe with large volume, one should require that $\epsilon\ll b_o$, $v_o\epsilon\gg 1$, $\sigma\ll \phi_o$ and $p_\phi\sigma\gg 1$. Denote by $\langle\hat{O}\rangle:=\frac{\langle \Psi|\hat{O}|\Psi\rangle}{\langle \Psi|\Psi\rangle}$ the expectation value of an operator $\hat{O}$ under the coherent states \eqref{eqn:shadow-grav}. By using the Poisson resummation on the sum over $n$ and the steepest descent approximation, the expectation value of each term of the gravitational Hamiltonian constraint operator $\hat{\cal H}^{k=-1}_{\rm grav}$ can be calculated, and thus the resulting expectation value of $\hat{\cal H}^{k=-1}_{\rm grav}$ can be obtained. For brevity, in the remainder of this paper, we will suppress the label $o$ appearing in $b_o, v_o$ and $\phi_o$. A straightforward calculation  reveals that (see Appendix \ref{appendix} for a derivation) \cite{Yang:2009fp,Assanioussi:2018hee,Assanioussi:2019iye}
\begin{widetext}
\begin{align}
 {\cal H}^{{\rm E},k=0}_{\rm grav,eff}&:=\langle\hat{\cal H}^{{\rm E},k=0}_{\rm grav}\rangle=\frac{3\hbar\gamma v}{4\sqrt{\Delta}}\left[\sin^2(2b)+O(\epsilon^2)\right]\left\{1+O(e^{-\pi^2/\epsilon^2})+O\left[1/(v\epsilon)^2\right]\right\},\\
 {\cal H}^{{\rm L},k=0}_{\rm grav,eff}&:=\langle\hat{\cal H}^{{\rm L},k=0}_{\rm grav}\rangle=\frac{3\hbar v}{32\gamma\sqrt{\Delta}}\left[\sin^2(4b)+O(\epsilon^2)\right]\left\{1+O(e^{-\pi^2/\epsilon^2})+O\left[1/(v\epsilon)^2\right]\right\},\\
 {\cal H}^{\Gamma,k=-1}_{\rm grav, eff}&:=\langle\hat{\cal H}^{\Gamma,k=-1}_{\rm grav}\rangle=\frac{3\left(\gamma\sqrt{\Delta}\,\hbar\right)^{\frac13}V_o^{\frac23}}{4\left(2\pi G\right)^{\frac23}}{v}^{\frac13}\left\{1+O(e^{-\pi^2/\epsilon^2})+O\left[1/(v\epsilon)^2\right]\right\}.
\end{align}
\end{widetext}
Hence, in the region with $\epsilon\ll b$, $v\epsilon\gg 1$, $\sigma\ll \phi$ and $p_\phi\sigma\gg 1$ the higher-order corrections can be omitted. In what follows, we focus on the leading terms. Hence the effective Hamiltonian constraint of gravitational part for the $k=-1$ model reads
\begin{align}
 {\cal H}^{k=-1}_{\rm grav,eff}&={\cal H}^{{\rm E},k=0}_{\rm grav,eff}-2(1+\gamma^2){\cal H}^{{\rm L},k=0}_{\rm grav,eff}+{\cal H}^{\Gamma,k=-1}_{\rm grav, eff}\notag\\
 &=-\frac{3\hbar v}{\gamma\sqrt{\Delta}}\left\{\frac14 \sin^2(2b)\left[1-(1+\gamma^2)\sin^2(2b)\right]\right.\notag\\
 &\hspace{3cm}\left.-\frac{\gamma^2\Delta}{4}\frac{V_o^{2/3}}{V^{2/3}}\right\}\notag\\
 &\equiv-\frac{3\hbar v}{\gamma\sqrt{\Delta}} g_{\rm eff}(b,v).
\end{align}
Taking into account the result for the scalar field in \cite{Yang:2009fp}, the total effective Hamiltonian constraint of the gravity coupled with a massless scalar field reads
\begin{align}\label{eq:eff-tot-Ham}
 {\cal H}^{k=-1}_{\rm tot,eff}&=-\frac{3\hbar v}{\gamma\sqrt{\Delta}} g_{\rm eff}(b,v)+\frac{p_\phi^2}{4\pi\gamma G\hbar\sqrt{\Delta}\,v}.
\end{align}
Before calculating the dynamics of this effective Hamiltonian, we clarify some subtle issues.  First, we expect here that the evolution of, saying, $\langle \hat v\rangle$ up to $O(\hbar)$ order coincides with the dynamics determined by the effective Hamiltonian constraint \eqref{eq:eff-tot-Ham}. In other words, if we compute the quantum dynamics of the coherent state \eqref{eqn:shadow-grav} and investigate the evolution of the expectation value of $\hat v$, we conjecture that the result coincides with the dynamics of $v$ obtained by solving the Hamilton's equation concerning the effective Hamiltonian \eqref{eq:eff-tot-Ham}. Second, as claimed before, we consider the region with  $\epsilon\ll b$ so that the higher-order corrections are omitted. However, as shown later, in the FRW phase of the evolution given by the effective Hamiltonian, $b$ approaches $0$ asymptotically. We thus obtain a tension that, on the one hand, we require $b\gg\epsilon$ but, on the other hand, $b$ goes to $0$ along the evolution. To resolve this tension, we still have to compute the quantum dynamics to see how the spread $\epsilon=1/\sqrt{2|\langle \hat v\rangle^2-\langle \hat v^2\rangle|}$ evolves. Even though the quantum dynamics, which will be left as our future work, has not been investigated yet, the previous results in the $k=0$ model \cite{Ashtekar:2006wn,Zhang:2019dgi} make us expect $\epsilon\sim 1/\langle\hat v\rangle$ which would make $\epsilon\ll b$ true along the evolution. Indeed, coherent states with the phase-space dependent spread have been considered in the regular LQC \cite{Taveras:2008ke}. Moreover, another approach to understand the effective dynamics is to apply the path integral formulation to study transition amplitude $A(v_f,\phi_f;v_i,\phi_i)=\langle v_f,\phi_f|v_i,\phi_i\rangle_{\rm phy}$ with $\langle\cdot|\cdot\rangle_{\rm phy}$ denoting the physical inner products \cite{Ashtekar:2010ve,Qin:2012gaa}. Since the Hamiltonian constraint operator $\hat{\cal H}^{k=-1}_{\rm grav}$ is the same as that of the $k=0$ model up to the term $\hat{\cal H}^{\Gamma,k=-1}_{\rm grav,eff}$ which takes $|v\rangle$ as its eigenstate, one can simply generalize the results in \cite{Qin:2012gaa} to conclude that the classical path resulting from dynamics of the effective Hamiltonian dominates $A(v_f,\phi_f;v_i,\phi_i)$.} Finally, in the region with $b\rightarrow0$ and $v\gg  1$, it is easy to see that, as $b\rightarrow0$, $g_{\rm eff}(b,v)$ goes to $g(b,v)$, and thus the total effective Hamiltonian constraint ${\cal H}^{k=-1}_{\rm tot,eff}$ in Eq. \eqref{eq:eff-tot-Ham} reduces to its classical expression ${\cal H}_{\rm tot}^{k=-1}$ in Eq. \eqref{eq:class-tot-Ham}. Hence the new alternative quantum dynamics has the corrected classical limit.

\begin{figure}
  \includegraphics[width=0.9\columnwidth]{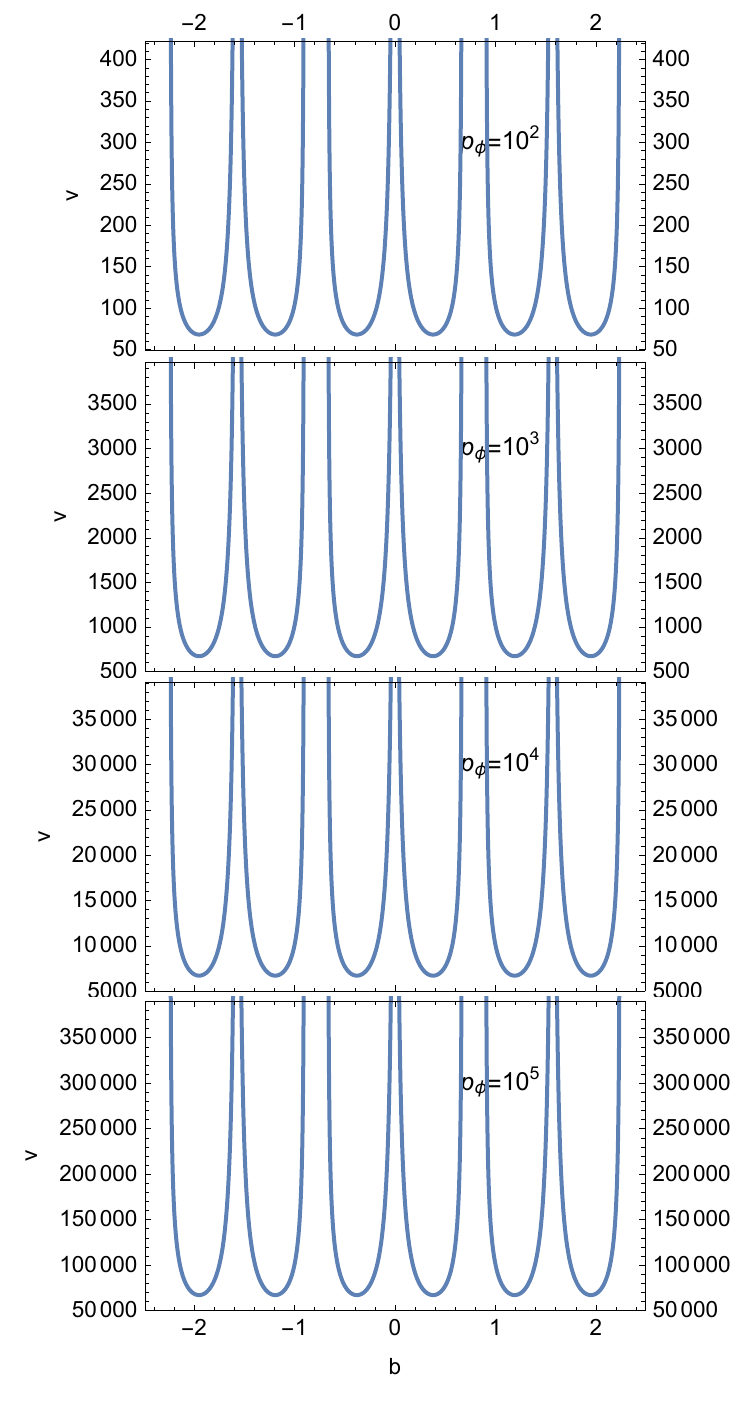}
  \caption{Plots of $v$ with respect to $b$ determined by the total Hamiltonian constraint equation ${\cal H}^{k=-1}_{\rm tot,eff}=0$ for different values of $p_\phi$, with $V_0=G=\hbar=1$ and $\gamma=0.2375$.}
  \label{fig:v-b}
\end{figure}

\begin{figure}[h]
  \includegraphics[width=0.9\columnwidth]{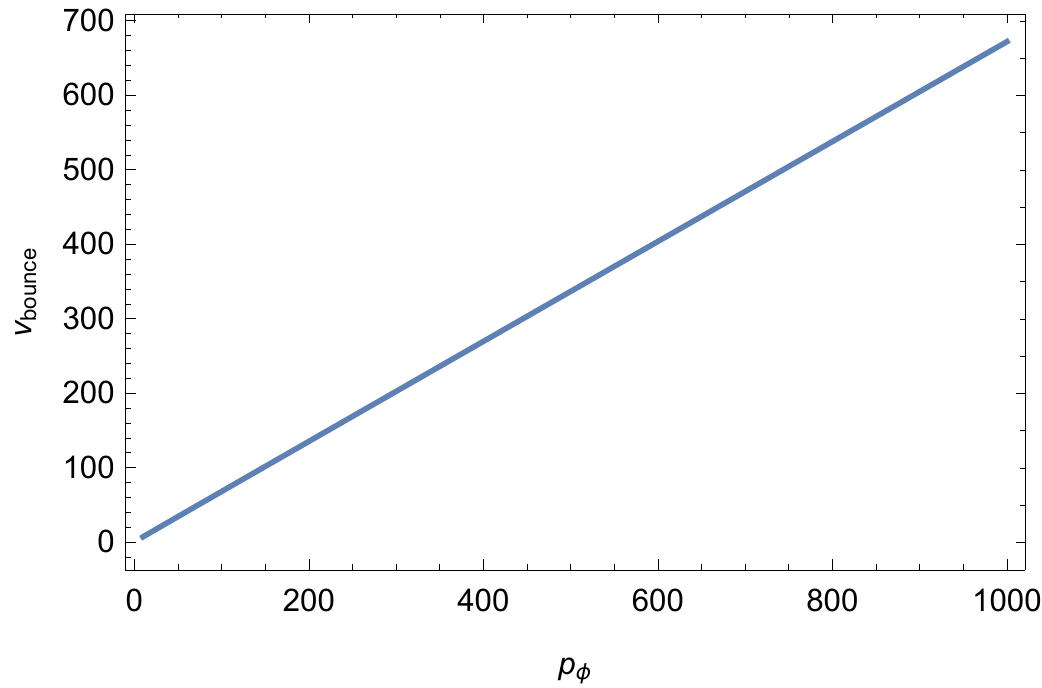}
  \caption{Plot of $v_{\rm bounce}$ with respect to $p_\phi$, with $V_0=G=\hbar=1$ and $\gamma=0.2375$.}
  \label{fig:v-bounce}
\end{figure}

It is easy to see that, in the effective theory, $p_\phi$ is a constant of motion due to $\dot{p}_\phi=\{p_\phi,{\cal H}^{k=-1}_{\rm tot,eff}\}=0$, and $\phi$ can be regarded as an internal clock because of $\dot{\phi}>0$, similar to the classical theory. The effective Hamiltonian constraint equation
\begin{align}\label{eq:eff-constraint-equation}
 {\cal H}^{k=-1}_{\rm tot,eff}=0
\end{align}
can determine the evolutions of $v$ with respect to $b$ for some given $p_\phi$, which is plotted in Fig. \ref{fig:v-b}.
Figure \ref{fig:v-b} depicts that $v=0$ can never be a solution to Eq. \eqref{eq:eff-constraint-equation}. It indicates that the classical singularity at $v=0$ can be avoided in the effective theory. By Eq. \eqref{eq:eff-constraint-equation}, the matter density can be expressed as
\begin{align}\label{rho-eff}
 \rho_\phi(v)&=\frac{p_\phi^2}{2V^2}=-\frac{{\cal H}^{k=-1}_{\rm grav,eff}}{V}\notag\\
 &=\frac{3}{2\pi G \gamma^2\Delta} g_{\rm eff}(b,v)\equiv\rho_\phi^{\rm eff}(b,v).
\end{align}
The effective Hubble parameter is determined by the total effective Hamiltonian constraint \eqref{eq:eff-tot-Ham}, and reads
\begin{align}\label{eq:eff-Hubble-expression-minus}
H^2_{{\rm eff},k=-1}&=\left(\frac{\dot{v}}{3v}\right)^2=\left(\frac{\left\{v,{\cal H}^{k=-1}_{\rm tot,eff}\right\}}{3v}\right)^2\notag\\
&=\frac{1}{\gamma^2\Delta}[g_{\rm eff}'(b,v)]^2,
\end{align}
where $'$ denotes the first-order derivative with respect to $b$. In what follows, we focus on the region $b\in[0,\pi/4]$ where we live. A bounce appears when $H^2_{{\rm eff},k=-1}=0$, i.e.,
\begin{align}
g_{\rm eff}'(b,v)=0\quad\Leftrightarrow\quad b=\frac12\arcsin\sqrt{\frac{1}{2(1+\gamma^2)}},
\end{align}
at which the energy density $\rho_\phi$ takes the maximal value, namely the critical energy density, as
\begin{align}\label{rho-critical}
\rho^{k=-1}_{\rm crit}&=\rho_F-\frac{3}{8\pi G}\frac{V_o^{2/3}}{V^{2/3}}=\rho_F-\frac{3}{8\pi G}\frac{1}{a^2},
\end{align}
where
\begin{align}
 \rho_F&:=\frac{3}{32\pi G\gamma^2(1+\gamma^2)\Delta}.
\end{align}
In comparison with the effective $k=0$ model proposed in \cite{Yang:2009fp} where the critical density is given by the first term of Eq. \eqref{rho-critical}, the effective $k=-1$ model contributes an additional term, the second term of Eq. \eqref{rho-critical}, to the critical energy density. It is worth mentioning that the critical energy density $\rho^{k=-1}_{\rm crit}$ in the effective $k=-1$ model depends on the value of $v$ (or the scale factor $a$) at the bounce point. The value $v_{\rm bounce}$ of $v$ at the bounce point can be determined by
\begin{align}
 \rho_\phi(v_{\rm bounce})=\rho^{k=-1}_{\rm crit}(v_{\rm bounce}).
\end{align}
In Fig. \ref{fig:v-bounce}, the value $v_{\rm bounce}$ as a function of $p_\phi$ is plotted. It is shown that as $p_\phi$ increase, the value $v_{\rm bounce}$ increase monotonically. Moreover the condition $\rho_\phi(v_{\rm bounce})\geq0$ implies that
\begin{align}
 a\geq\sqrt{4\gamma^2(1+\gamma^2)\Delta},
\end{align}
and thus the effective theory predicts that the scale factor $a$ is bounced below $a_{\rm min}=\sqrt{4\gamma^2(1+\gamma^2)\Delta}$.

Solving the total effective Hamiltonian constraint equation ${\cal H}^{k=-1}_{\rm tot,eff}=0$ for $b$ yields
\begin{align}\label{b-rho-relation-minus}
b=
\begin{cases}
b_{\rm I}\equiv\frac12\arcsin\sqrt{\frac{1-\sqrt{1-\frac{\rho_\phi+\frac{3}{8\pi G}\frac{1}{a^2}}{\rho_F}}}{2(1+\gamma^2)}}\\
b_{\rm II}\equiv\frac12\arcsin\sqrt{\frac{1+\sqrt{1-\frac{\rho_\phi+\frac{3}{8\pi G}\frac{1}{a^2}}{\rho_F}}}{2(1+\gamma^2)}}
\end{cases}.
\end{align}
Hence there exists two types of classical universe, namely the type-I universe and the type-II universe. The two universes are connected by a quantum bounce. The effective Hubble parameter is determined by
\begin{widetext}
\begin{align}\label{eq:eff-Friedmann-eq}
 H_{{\rm eff},k=-1,{\rm I/II}}^2&=\left.\left(\frac{\dot{v}}{3v}\right)^2\right|_{b=b_{\rm I/II}}\notag\\
 &=\frac{8\pi G}{3} \frac{\rho_F}{1+\gamma^2}\left(1-\frac{\rho_\phi+\frac{3}{8\pi G}\frac{1}{a^2}}{\rho_F}\right) \left(1\mp\sqrt{1-\frac{\rho_\phi+\frac{3}{8\pi G}\frac{1}{a^2}}{\rho_F}}\right)\left(1+2 \gamma^2\pm\sqrt{1-\frac{\rho_\phi+\frac{3}{8\pi G}\frac{1}{a^2}}{\rho_F}}\right),
\end{align}
which can be expressed as the following more convenient forms
\begin{align}\label{eq:H2-eff-minus}
  H_{{\rm eff},k=-1,{\rm I}}^2
  &=\frac{8\pi G}{3}\left(\rho_\phi+\frac{3}{8\pi G}\frac{1}{a^2}\right)\left(1-\frac{\rho_\phi+\frac{3}{8\pi G}\frac{1}{a^2}}{\rho_F}\right)\left[1+\frac{\gamma^2}{1+\gamma^2}
  \left(\frac{\sqrt{\frac{\rho_\phi+\frac{3}{8\pi G}\frac{1}{a^2}}{\rho_F}}}{1+\sqrt{1-\frac{\rho_\phi+\frac{3}{8\pi G}\frac{1}{a^2}}{\rho_F}}}\right)^2\right],
\end{align}
and
\begin{align}\label{eq:H2-eff-plus}
 H_{{\rm eff},k=-1,{\rm II}}^2&=\frac{8\pi G}{3}\rho_{\Lambda_{\rm eff}}\left(1-\frac{\rho_\phi+\frac{3}{8\pi G}\frac{1}{a^2}}{\rho_F}\right)\left[1+\frac{1-2\gamma^2+\sqrt{1-\frac{\rho_\phi+\frac{3}{8\pi G}\frac{1}{a^2}}{\rho_F}}}{4\gamma^2\left(1+\sqrt{1-\frac{\rho_\phi+\frac{3}{8\pi G}\frac{1}{a^2}}{\rho_F}}\right)}\frac{\rho_\phi+\frac{3}{8\pi G}\frac{1}{a^2}}{\rho_F}\right],
\end{align}
\end{widetext}
where
\begin{align}
 \rho_{\Lambda_{\rm eff}}:=\frac{\Lambda_{\rm eff}}{8\pi G},\qquad{\text{with}}\quad\Lambda_{\rm eff}:=\frac{3}{(1+\gamma^2)^2\Delta}.
\end{align}

Now let us study the asymptotic behavior of the effective dynamics at the large $v$ limit. For $v\rightarrow\infty$, the matter density $\rho_\phi(v)$ in Eq. \eqref{rho-eff} goes to zero, and thus
\begin{align}
\sin^2(2b)\left[1-(1+\gamma^2)\sin^2(2b)\right]\rightarrow0,
\end{align}
which implies
\begin{align}
b\rightarrow b_0=
\begin{cases}
b_{\rm I, c}\equiv0,\\
b_{\rm II, c}\equiv\frac12\arcsin\left(\frac{1}{\sqrt{1+\gamma^2}}\right).
\end{cases}
\end{align}
Expanding ${\cal H}^{k=-1}_{\rm tot,eff}$ at $b_0$ up to the second order yields the classical behavior of the effective Hamiltonian constraint as
\begin{align}
{\cal H}^{k=-1}_{\rm tot,eff}\;\rightarrow\;&-\frac{3\hbar v}{\gamma\sqrt{\Delta}}\Big[g_{\rm eff}(b_0,v)+g_{\rm eff}'(b_0)(b-b_0)\notag\\
&\hspace{1cm}+\frac{g_{\rm eff}''(b_0)}{2}(b-b_0)^2\Big]+\frac{p_\phi^2}{4\pi \gamma G\hbar\sqrt{\Delta}\,v}\,,
\end{align}
where $''$ denotes the second-order derivative with respective to $b$, and
\begin{align}
g_{\rm eff}'(b_0)=
\begin{cases}
0,& b_0=b_{\rm I, c},\\
-\frac{\gamma}{1+\gamma^2},& b_0=b_{\rm II, c},
\end{cases}\\
g_{\rm eff}''(b_0)=
\begin{cases}
2,& b_0=b_{\rm I, c},\\
\frac{2-10\gamma^2}{1+\gamma^2},& b_0=b_{\rm II, c}.
\end{cases}
\end{align}
Plugging these asymptotic expressions into Eq. \eqref{eq:eff-Hubble-expression-minus}, we have
\begin{widetext}
\begin{align}\label{eq:H_eff2-asymptotic}
 H_{{\rm eff},k=-1}^2&\rightarrow\frac{8\pi G}{3}\left(\frac{g''_{\rm eff}(b_0)}{2}\,\rho_\phi+\frac{3\left[g'_{\rm eff}(b_0)\right]^2}{8\pi G\gamma^2\Delta}+\frac{3g''_{\rm eff}(b_0)}{16\pi G}\frac{V_0^{2/3}}{V^{2/3}}\right)
 =\frac{8\pi G}{3}\left[\frac{g''_{\rm eff}(b_0)}{2}\,\left(\rho_\phi+\frac{3}{8\pi G}\frac{1}{a^2}\right)+\frac{3\left[g'_{\rm eff}(b_0)\right]^2}{8\pi G\gamma^2\Delta}\right]\notag\\
 &=
 \begin{cases}
 \frac{8\pi G}{3}\rho_\phi+\frac{1}{a^2},& b_0=b_{\rm I, c},\\
 \frac{8\pi G}{3}\left(\frac{1-5\gamma^2}{1+\gamma^2}\right)\rho_\phi+\left(\frac{1-5\gamma^2}{1+\gamma^2}\right)\frac{1}{a^2}+\frac{\Lambda_{\rm eff}}{3},& b_0=b_{\rm II, c}.
 \end{cases}
\end{align}
\end{widetext}
The above asymptotic behavior \eqref{eq:H_eff2-asymptotic} of the effective Hubble parameter can be also obtained directly from Eqs. \eqref{eq:H2-eff-minus} and \eqref{eq:H2-eff-plus}. Equation \eqref{eq:H_eff2-asymptotic} implies that the type-II universe is an asymptotic de Sitter universe with a positive effective cosmological constant $\Lambda_{\rm eff}$. Therefore the asymptotical $k=-1$ FRW universe (the type-I universe) will be bounced to an asymptotic de Sitter universe (the type-II universe) coupled to a scalar field.

\begin{figure}
  \includegraphics[width=0.9\columnwidth]{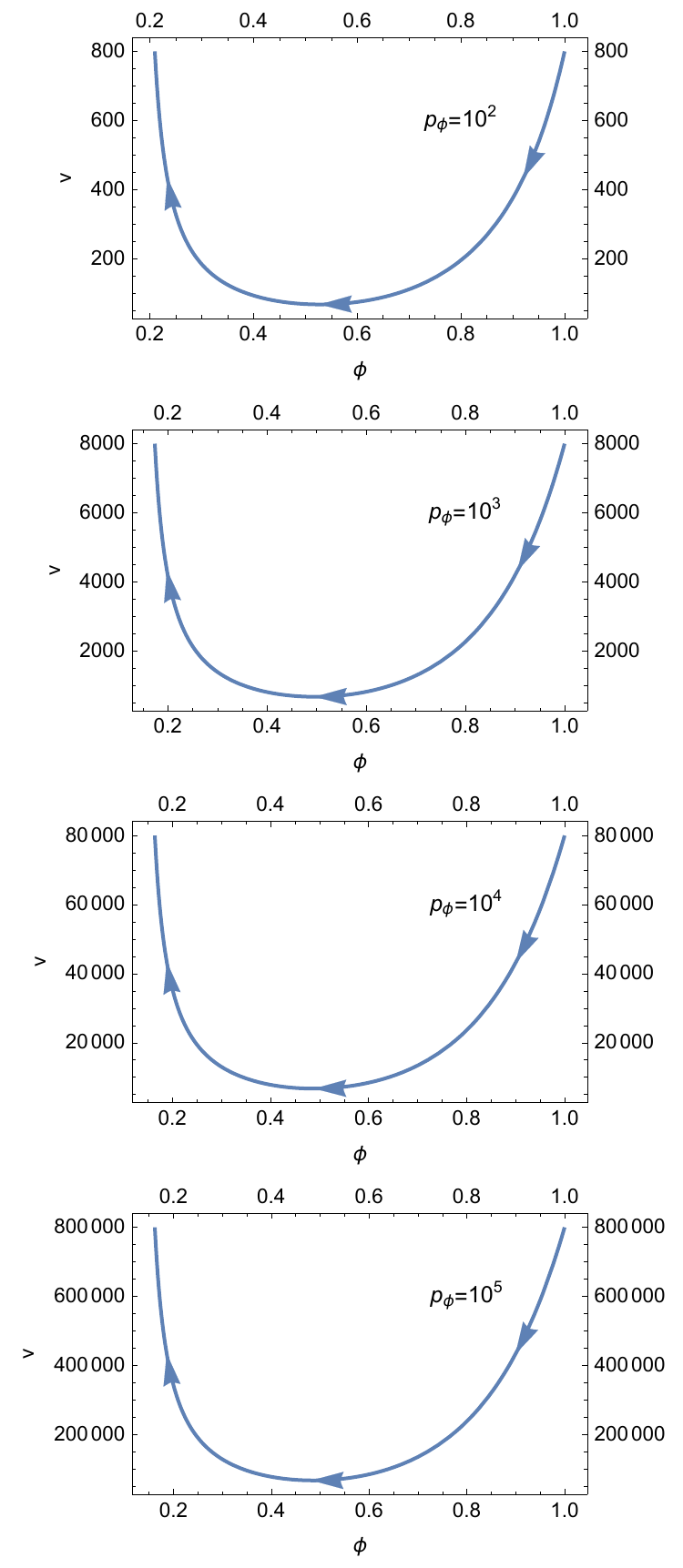}
  \caption{Plots of the effective dynamical evolution of $v$ with respect to $\phi$, in reverse direction of the cosmological time, determined by the effective Hamiltonian constraint for different values of $p_\phi$ with initial data $v(b_{\rm start},p_\phi)=8p_\phi$ and $\phi(b_{\rm start},p_\phi)=1$, with $V_0=G=\hbar=1$ and $\gamma=0.2375$.}
  \label{fig:v-phi}
\end{figure}

We now numerically study the the effective dynamical evolution of $v$ with $\phi$. To this end, we can firstly solve the effective Hamiltonian constraint equation ${\cal H}^{k=-1}_{\rm tot,eff}=0$ to yield
\begin{align}\label{eq:v-b}
 v=v(b,p_\phi).
\end{align}
Secondly, we consider the evolution equation of $\phi$ with respect to $b$, namely
\begin{align}\label{eq:phi-b-equation}
 \frac{{\rm d}\phi}{{\rm d}b}=\frac{\{\phi,{\cal H}^{k=-1}_{\rm tot,eff}\}}{\{b,{\cal H}^{k=-1}_{\rm tot,eff}\}}= f(b,p_\phi),
\end{align}
where Eq. \eqref{eq:v-b} was inserted in the second step. Solving Eq. \eqref{eq:phi-b-equation} yields
\begin{align}\label{eq:phi-b}
 \phi=\phi(b,p_\phi).
\end{align}
By combining Eq. \eqref{eq:v-b} with Eq. \eqref{eq:phi-b} and then eliminating $b$, we arrive at $v=v(\phi,p_\phi)$. In Fig. \ref{fig:v-phi}, the effective dynamical evolution of $v$ with respect to $\phi$ for given $p_\phi$ is plotted. It indicates that an asymmetric bounce appears in the backward evolution of the universe sourced by a massless scalar field $\phi$, and the classical big-bang singularity is resolved.

\section{Summary}\label{section5}

The quantization ambiguities often exist in constructing the gravitational Hamiltonian constraint operator of full LQG as well as of LQC. It has been shown that in LQC different quantizations of the gravitational Hamiltonian constraint may lead to different quantum dynamics, and thus affect the fate of the universe. Hence the study of the quantization ambiguities of the gravitational Hamiltonian constraint plays an important role in the quantum dynamics of LQC. In present paper, we have studied an alternative quantization of the gravitational Hamiltonian constraint in the $k=-1$ model of LQC closely following that in the $k=0$ model proposed in \cite{Yang:2009fp}, mimicking the treatment of full LQG.

Classically, the connection $A^i_a$ in the $k=-1$ model takes the nondiagonal expression \eqref{eq:A-E-forms-A} on the left-invariant one-forms ${}^o\!\omega^i_a$, while it takes the diagonal form on the left- and right-invariant one-forms ${}^o\!\omega^i_a$ in the $k=0$ model. In the $k=-1$ model, the nondiagonal expression of $A^i_a$ leads to the complicated forms of the resulting holonomies of the connection. Instead, one often considers the holonomies of the extrinsic curvature $K^i_a$ multiplied by $\gamma$ in the $k=-1$ model, and thus they have the same expressions as the holonomies of the connection $A^i_a=\gamma K^i_a$ in the $k=0$ model. Hence both the $k=0$ model and the $k=-1$ model have the same Hilbert space ${\mathscr H}^{\rm gr}_{\rm kin}=L^2(\mathbb{R}_{\mathrm{Bohr}},{\rm d}\mu_{\mathrm{Bohr}})$.

To study the quantum dynamics of the $k=0,-1$ models in the framework of LQC, one needs to promote the gravitational Hamiltonian constraint into a well-defined operator in ${\mathscr H}^{\rm gr}_{\rm kin}$. At the classical level, the gravitational Hamiltonian constraint of the $k=-1$ model can be expressed as the three terms in Eq. \eqref{classical-Hamiltonian-minus}. On one hand, the former two terms, $\tilde{\cal H}^{{\rm E},k=0}_{\rm grav}$ and $\tilde{\cal H}^{{\rm L},k=0}$, in Eq. \eqref{classical-Hamiltonian-minus} have the same expressions as the Euclidean and Lorentzian terms, ${\cal H}^{{\rm E},k=0}_{\rm grav}$ and ${\cal H}^{{\rm L},k=0}$, in the $k=0$ model, respectively, although involving different fiducial one-forms ${}^o\!\omega^i_a$ (the only left-invariant one-forms v.s. the left- and right-invariant one-forms). From the symmetry-reduced expressions in Eqs. \eqref{eq:Euclidean-term} and \eqref{eq:Lorentz-term}, the former two terms are proportional to each other, and thus one can firstly combine the two terms and then quantize the first and third terms to obtain the gravitational Hamiltonian constraint \cite{Vandersloot:2006ws}. On the other hand, from the viewpoint of full LQG, the sum of both the first and last terms forms the Euclidean term, while the middle term represents the Lorentzian term. Hence, alternatively, we can quantize the Euclidean term and the Lorentzian term respectively, mimicking the treatment in full LQG. We have shown that the former two terms $\tilde{\cal H}^{{\rm E},k=0}_{\rm grav}$ and $\tilde{\cal H}^{{\rm L},k=0}$ in Eq. \eqref{classical-Hamiltonian-minus} can be quantized as the operators, $\hat{\cal H}^{{\rm E},k=0}_{\rm grav}$ and $\hat{\cal H}^{{\rm L},k=0}_{\rm grav}$ in Eqs. \eqref{eq:k0-Eculdian-op} and \eqref{eq:k0-Lorentzian-op}, corresponding precisely to the Euclidean and Lorentzian Hamiltonian operators of the $k=0$ model proposed in \cite{Yang:2009fp}, while the third term has been quantized as $\hat{\cal H}^{\Gamma,k=-1}_{\rm grav}$ in Eq. \eqref{eq:k0-Gamma-op}. The resulting gravitational Hamiltonian constraint operator $\hat{\cal H}^{k=-1}_{\rm grav}$ is symmetric, which has the action \eqref{eq:action-k-1model} on $|v\rangle$. Moreover, we have shown that the new quantum dynamics determined by the alternative Hamiltonian constraint operator $ \hat{\cal H}_{\rm tot}^{k=-1}$ in Eq. \eqref{eq:quantum-equation} for the $k=-1$ model coupled to a massless scalar field has the corrected classical limit, and obtained its effective Hamiltonian constraint \eqref{eq:eff-tot-Ham}, by semi-classical analysis. The effective Friedmann equation for the $k=-1$ model was derived in Eq. \eqref{eq:eff-Friedmann-eq}, which shows that it has two branches \eqref{eq:H2-eff-minus} and \eqref{eq:H2-eff-plus} relating to two types of universes, similar to the $k=0$ LQC proposed in \cite{Yang:2009fp}. It turns out that the asymptotical $k=-1$ FRW universe (the type-I universe) will be bounced to an asymptotic de Sitter universe (the type-II universe) coupled to a scalar field. Last but not least, by requiring the condition $\rho_\phi(v_{\rm bounce})\geq0$, the effective theory predicts that the scale factor $a$ is bounced below $a_{\rm min}=\sqrt{4\gamma^2(1+\gamma^2)\Delta}$, which is different from that in previous $k=-1$ LQC model \cite{Vandersloot:2006ws}.

So far, the Thiemann's trick for regularizing the gravitational Hamiltonian constraint in full LQG, by treating the Euclidean term and the Lorentzian term independently, has been successfully applied to the $k=0,-1$ LQC models. However, to our knowledge, a similar treatment for the $k=+1$ model in the framework of LQC has not been carried out. In spite of the Thiemann-regularization of the Hamiltonian constraint on the hyperspherical lattice for the $k = +1$ model has been studied from the viewpoint of full LQG in \cite{Liegener:2019out}. The expectation value of the Hamiltonian constraint under certain coherent states was computed, and an effective Hamiltonian constraint was obtained in the $\mu_o$ scheme rather than the $\bar{\mu}$ scheme \cite{Liegener:2019out}. By numerical simulations of the dynamical evolution, an asymmetric bounce replacing the classical big bang was also obtained in the model \cite{Liegener:2019out}.

It should be noted that there are still many aspects of the loop quantum $k=-1$ model that deserves further investigation. Recently, some works have focused on the relation between LQG and the $k=0$ LQC by calculating the expectation value of the Hamiltonian in LQG under certain coherent state peaked at some point in the classical phase space \cite{Dapor:2017rwv,Han:2019vpw,Han:2019feb,Liegener:2020dcg,Zhang:2020mld,Zhang:2021qul}. How to generate these works to the $k=-1$ case will be interesting. Moreover, except the alternative regularization from the Thiemann's trick adopted in the present paper following directly that in the $k=0$ case, the other alternative regularizations employed in the $k=0$ model \cite{Liegener:2019zgw,Yang:2019ujs,Yang:2020eby,Zhang:2021zfp} can also be in principle extended to the $k=-1$ model.

\begin{acknowledgements}
This work is supported in part by NSFC Grants No. 12165005, No. 11961131013, No.12275087, No. 11775082, and ``the Fundamental Research Funds for the Central Universities''.
\end{acknowledgements}

\begin{appendix}
\section{Derivation of the expectation values}\label{appendix}
In this appendix, we present the calculations of expectation values of the Hamiltonian constraint operator on the coherent states. To this end, let us firstly consider an operator $\hat{H}$ with the action on $|v\rangle$
\begin{align}
 \hat{H}|v\rangle&=f_+(v)|v+l\rangle+f_0(v)|v\rangle+f_-(v)|v-l\rangle,
\end{align}
where $l$ is a positive even number, and
\begin{align}\label{eq:coeff-behavior}
 f_-(v)=f_+(v-l).
\end{align}
Then the expectation value of $\hat{H}$ on the coherent states \eqref{eqn:shadow-grav} reads
\begin{widetext}
\begin{align}
 \langle\Psi_{\rm grav}|\hat{H}|\Psi_{\rm grav}\rangle&=\sum_{n,n'\in\mathbb{Z}}e^{-\frac{\epsilon^2}{2}(n'-v_o)^2}e^{{\rm i}b_o(n'-v_o)}e^{-\frac{\epsilon^2}{2}(n-v_o)^2}e^{-{\rm i}b_o(n-v_o)}\notag\\
 &\hspace{2cm}\times\left[f_+(n)\langle n'|n+l\rangle+f_-(n)\langle n'|n-l\rangle+f_0(n)\langle n'|n\rangle\right]\notag\\
 &=\sum_{n\in\mathbb{Z}}e^{-\frac{\epsilon^2}{2}(n+l-v_o)^2}e^{{\rm i}b_o(n+l-v_o)}e^{-\frac{\epsilon^2}{2}(n-v_o)^2}e^{-{\rm i}b_o(n-v_o)}f_+(n)\notag\\
 &\quad+\sum_{n\in\mathbb{Z}}e^{-\frac{\epsilon^2}{2}(n-l-v_o)^2}e^{{\rm i}b_o(n-l-v_o)}e^{-\frac{\epsilon^2}{2}(n-v_o)^2}e^{-{\rm i}b_o(n-v_o)}f_-(n)\notag\\
 &\quad+\sum_{n\in\mathbb{Z}}e^{-\epsilon^2(n-v_o)^2}f_0(n)\notag\\
 &=\sum_{n\in\mathbb{Z}}e^{-\frac{\epsilon^2}{2}(n+l-v_o)^2}e^{{\rm i}b_o(n+l-v_o)}e^{-\frac{\epsilon^2}{2}(n-v_o)^2}e^{-{\rm i}b_o(n-v_o)}f_+(n)\notag\\
 &\quad+\sum_{n'\in\mathbb{Z}}e^{-\frac{\epsilon^2}{2}(n'-v_o)^2}e^{{\rm i}b_o(n'-v_o)}e^{-\frac{\epsilon^2}{2}(n'+l-v_o)^2}e^{-{\rm i}b_o(n'+l-v_o)}f_+(n')\notag\\
 &\quad+\sum_{n\in\mathbb{Z}}e^{-\epsilon^2(n-v_o)^2}f_0(n)\notag\\
 &=(e^{{\rm i}lb_o}+e^{-{\rm i}lb_o})\sum_{n'\in\mathbb{Z}}e^{-\frac{\epsilon^2}{2}\left[(n'-v_o)^2+(n'+l-v_o)^2\right]}f_+(n')+\sum_{n\in\mathbb{Z}}e^{-\epsilon^2(n-v_o)^2}f_0(n)\notag\\
 &=2\cos(lb_o)e^{-\frac{l^2}{4}\epsilon^2}\sum_{n\in\mathbb{Z}}e^{-\epsilon^2\left(n-v_o\right)^2}f_+\left(n-\frac{l}{2}\right)+\sum_{n\in\mathbb{Z}}e^{-\epsilon^2(n-v_o)^2}f_0(n)\notag\\
 &=\left[\sin^2\left(\frac{l}{2}b_o\right)-\frac12\right]e^{-\frac{l^2}{4}\epsilon^2}\left[-4\sum_{n\in\mathbb{Z}}e^{-\epsilon^2\left(n-v_o\right)^2}f_+\left(n-\frac{l}{2}\right)\right]+\frac12\left[2\sum_{n\in\mathbb{Z}}e^{-\epsilon^2(n-v_o)^2}f_0(n)\right],
\end{align}
where in the third step we have used Eq. \eqref{eq:coeff-behavior} and relabeled $n-l$ by $n'$. Applying the Possion resummation formula and the steepest decent method, for an arbitrary analytic function $g(n)$, one has \cite{Ashtekar:2003hd}
\begin{align}
 \sum_{n\in\mathbb{Z}}e^{-\epsilon^2\left(n-v_o\right)^2}g(n)=\frac{\sqrt{\pi}}{\epsilon}g(v_o)\left\{1+O(e^{-\pi^2/\epsilon^2})+O\left[1/(v_o\epsilon)^2\right]\right\}.
\end{align}
Then one gets
\begin{align}
 \langle \Psi_{\rm grav}|\Psi_{\rm grav}\rangle&=\sum_{n,n'\in\mathbb{Z}}e^{-\frac{\epsilon^2}{2}(n'-v_o)^2}e^{{\rm i}b_o(n'-v_o)}e^{-\frac{\epsilon^2}{2}(n-v_o)^2}e^{-{\rm i}b_o(n-v_o)}\langle n'|n\rangle\notag\\
 &=\sum_{n\in\mathbb{Z}}e^{\epsilon^2(n-v_o)^2}\notag\\
 &=\frac{\sqrt{\pi}}{\epsilon}\left\{1+O(e^{-\pi^2/\epsilon^2})+O\left[1/(v_o\epsilon)^2\right]\right\},
\end{align}
and
\begin{align}
 \langle\Psi_{\rm grav}|\hat{H}|\Psi_{\rm grav}\rangle&=\frac{\sqrt{\pi}}{\epsilon}\left\{e^{-\frac{l^2}{4}\epsilon^2}\left[\sin^2\left(\frac{l}{2}b_o\right)-\frac12\right]\left[-4f_+\left(v_o-\frac{l}{2}\right)\right]+\frac12\left[2f_0(v_o)\right]\right\}\left\{1+O(e^{-\pi^2/\epsilon^2})+O\left[1/(v_o\epsilon)^2\right]\right\},
\end{align}
for analytic functions $f_+(y)$ and $f_0(y)$. If $f_+(y)$ [or $f_0(y)$] is not analytic function which is the case under considerations due to the involved absolute value, one can replace it with its analytic extention $\bar{f}_+(y)$ [or $\bar{f}_0(y)$] (for example, omitting the absolute value symbol). It turns out that the error $\sum_n\exp(-\epsilon^2(n-v_o))[f_+(n)-\bar{f}_+(n)]$ can be shown to be the order $O\left(e^{-v_o^2\epsilon^2}\right)$, which is negligible compared to the corrections derived above \cite{Ashtekar:2003hd}. Hence the resulting normalized expectation value of $\hat{H}$ reads
\begin{align}\label{eq:H-expectation}
\langle\hat{H}\rangle&=\frac{\langle\Psi|\hat{H}|\Psi\rangle}{\langle \Psi|\Psi\rangle}=\frac{\langle\Psi_{\rm grav}|\hat{H}|\Psi_{\rm grav}\rangle}{\langle \Psi_{\rm grav}|\Psi_{\rm grav}\rangle}\notag\\
&=\left\{e^{-\frac{l^2}{4}\epsilon^2}\left[\sin^2\left(\frac{l}{2}b_o\right)-\frac12\right]\left[-4f_+\left(v_o-\frac{l}{2}\right)\right]+\frac12\left[2f_0(v_o)\right]\right\}\left\{1+O(e^{-\pi^2/\epsilon^2})+O\left[1/(v_o\epsilon)^2\right]\right\}.
\end{align}

Now we turn to the three parts of the gravitational Hamiltonian constraint operator $\hat{\cal H}^{k=-1}_{\rm grav}$.  Applying the result in Eq. \eqref{eq:H-expectation} to the first two terms $\hat{\cal H}^{{\rm E},k=0}_{\rm grav}$ and $\hat{\cal H}^{{\rm L},k=0}_{\rm grav}$ with $l=4$ and $l=8$, respectively, we have
\begin{align}
 \langle\hat{\cal H}^{{\rm E},k=0}_{\rm grav}\rangle&=\frac{3\hbar\gamma v_o}{4\sqrt{\Delta}}\left[\sin^2\left(2b_o\right)e^{-4\epsilon^2}+\frac12\left(1-e^{-4\epsilon^2}\right)\right]\left\{1+O(e^{-\pi^2/\epsilon^2})+O\left[1/(v_o\epsilon)^2\right]\right\}\notag\\
 &=\frac{3\hbar\gamma v_o}{4\sqrt{\Delta}}\left[\sin^2\left(2b_o\right)+O(\epsilon^2)\right]\left\{1+O(e^{-\pi^2/\epsilon^2})+O\left[1/(v_o\epsilon)^2\right]\right\},\\
 \langle\hat{\cal H}^{{\rm L},k=0}_{\rm grav}\rangle
 &=\frac{3\hbar v_o}{32\gamma\sqrt{\Delta}}\left[\sin^2\left(4b_o\right)e^{-16\epsilon^2}+\frac12\left(1-e^{-16\epsilon^2}\right)\right]\left\{1+O(e^{-\pi^2/\epsilon^2})+O\left[1/(v_o\epsilon)^2\right]\right\}\notag\\
 &=\frac{3\hbar v_o}{32\gamma\sqrt{\Delta}}\left[\sin^2\left(4b_o\right)+O(\epsilon^2)\right]\left\{1+O(e^{-\pi^2/\epsilon^2})+O\left[1/(v_o\epsilon)^2\right]\right\},
\end{align}
where we have used the results
\begin{align}
-4\bar{E}_+(v_o-2)&=\frac{3\hbar\gamma}{4\sqrt{\Delta}}v_o,\qquad 2\bar{E}_0(v_o)=\frac{3\hbar\gamma}{4\sqrt{\Delta}}v_o,\\
-4\bar{L}_+(v_o-4)&=\frac{3\hbar}{32\gamma\sqrt{\Delta}}v_o,\qquad 2\bar{L}_0(v_o)=\frac{3\hbar}{32\gamma\sqrt{\Delta}}v_o.
\end{align}
To get the expectation value $\langle\hat{\cal H}^{\Gamma,k=-1}_{\rm grav}\rangle$ from Eq. \eqref{eq:H-expectation}, we drop the term involving $f_+(v)$ since $\hat{\cal H}^{\Gamma,k=-1}_{\rm grav}$ has $|v\rangle$ as its eigenstate, and obtain
\begin{align}
\langle\hat{\cal H}^{\Gamma,k=-1}_{\rm grav}\rangle&=\frac{3\left(\gamma\sqrt{\Delta}\,\hbar\right)^{\frac13}V_o^{\frac23}}{4\left(2\pi G\right)^{\frac23}}v_o^{\frac13}\left\{1+O(e^{-\pi^2/\epsilon^2})+O\left[1/(v_o\epsilon)^2\right]\right\},
\end{align}
where we have used
\begin{align}
\bar{\Gamma}(v_o)&=\frac{3\left(\gamma\sqrt{\Delta}\,\hbar\right)^{\frac13}V_o^{\frac23}}{4\left(2\pi G\right)^{\frac23}}v_o^{\frac13}.
\end{align}
\end{widetext}
\end{appendix}

\clearpage

%


\begin{thebibliography}{59}%

\bibitem{Rovelli:2004tv}
C.~Rovelli, \href{http://dx.doi.org/10.1017/CBO9780511755804}{{\em {Quantum
  Gravity}}} (Cambridge University Press, Cambridge, England, 2004).

\bibitem{Thiemann:2007pyv}
T.~Thiemann, \href{http://dx.doi.org/10.1017/CBO9780511755682}{{\em {Modern
  Canonical Quantum General Relativity}}} (Cambridge University Press,
  Cambridge, England, 2007).

\bibitem{Ashtekar:2004eh}
A.~Ashtekar and J.~Lewandowski, {Background independent quantum gravity: A
  status report}, \href{http://dx.doi.org/10.1088/0264-9381/21/15/R01}{Class.
  Quant. Grav. {\bfseries 21}, R53 (2004)}.

\bibitem{Han:2005km}
M.~Han, Y.~Ma, and W.~Huang, {Fundamental structure of loop quantum gravity},
  \href{http://dx.doi.org/10.1142/S0218271807010894}{Int. J. Mod. Phys. D
  {\bfseries 16}, 1397 (2007)}.

\bibitem{Rovelli:1994ge}
C.~Rovelli and L.~Smolin, {Discreteness of area and volume in quantum gravity},
  \href{http://dx.doi.org/10.1016/0550-3213(95)00150-Q}{Nucl. Phys. B
  {\bfseries 442}, 593 (1995)}.

\bibitem{Ashtekar:1996eg}
A.~Ashtekar and J.~Lewandowski, {Quantum theory of geometry: I. Area
  operators}, \href{http://dx.doi.org/10.1088/0264-9381/14/1A/006}{Class.
  Quant. Grav. {\bfseries 14}, A55 (1997)}.

\bibitem{Ashtekar:1997fb}
A.~Ashtekar and J.~Lewandowski, {Quantum theory of geometry II: Volume
  operators}, \href{http://dx.doi.org/10.4310/ATMP.1997.v1.n2.a8}{Adv. Theor.
  Math. Phys. {\bfseries 1}, 388 (1997)}.

\bibitem{Yang:2016kia}
J.~Yang and Y.~Ma, {New volume and inverse volume operators for loop quantum
  gravity}, \href{http://dx.doi.org/10.1103/PhysRevD.94.044003}{Phys. Rev. D
  {\bfseries 94}, 044003 (2016)}.

\bibitem{Thiemann:1996at}
T.~Thiemann, {A length operator for canonical quantum gravity},
  \href{http://dx.doi.org/10.1063/1.532445}{J. Math. Phys. {\bfseries 39}, 3372
  (1998)}.

\bibitem{Ma:2010fy}
Y.~Ma, C.~Soo, and J.~Yang, {New length operator for loop quantum gravity},
  \href{http://dx.doi.org/10.1103/PhysRevD.81.124026}{Phys. Rev. D {\bfseries
  81}, 124026 (2010)}.

\bibitem{Ashtekar:1997yu}
A.~Ashtekar, J.~Baez, A.~Corichi, and K.~Krasnov, {Quantum Geometry and Black
  Hole Entropy}, \href{http://dx.doi.org/10.1103/PhysRevLett.80.904}{Phys. Rev.
  Lett. {\bfseries 80}, 904 (1998)}.

\bibitem{Song:2020arr}
S.~Song, H.~Li, Y.~Ma, and C.~Zhang, {Entropy of black holes with arbitrary
  shapes in loop quantum gravity},
  \href{http://dx.doi.org/10.1007/s11433-021-1770-3}{Sci. China Phys. Mech.
  Astron. {\bfseries 64}, 120411 (2021)}.

\bibitem{Zhang:2011vi}
X.~Zhang and Y.~Ma, {Extension of Loop Quantum Gravity to $f(R)$ Theories},
  \href{http://dx.doi.org/10.1103/PhysRevLett.106.171301}{Phys. Rev. Lett.
  {\bfseries 106}, 171301 (2011)}.

\bibitem{Zhang:2011qq}
X.~Zhang and Y.~Ma, {Loop quantum $f(R)$ theories},
  \href{http://dx.doi.org/10.1103/PhysRevD.84.064040}{Phys. Rev. D {\bfseries
  84}, 064040 (2011)}.

\bibitem{Zhang:2011gn}
X.~Zhang and Y.~Ma, {Loop quantum Brans-Dicke theory},
  \href{http://dx.doi.org/10.1088/1742-6596/360/1/012055}{J. Phys. Conf. Ser.
  {\bfseries 360}, 012055 (2012)}.

\bibitem{Zhang:2011vg}
X.~Zhang and Y.~Ma, {Nonperturbative loop quantization of scalar-tensor
  theories of gravity},
  \href{http://dx.doi.org/10.1103/PhysRevD.84.104045}{Phys. Rev. D {\bfseries
  84}, 104045 (2011)}.

\bibitem{Bodendorfer:2011nv}
N.~Bodendorfer, T.~Thiemann, and A.~Thurn, {New variables for classical and
  quantum gravity in all dimensions: I. Hamiltonian Analysis},
  \href{http://dx.doi.org/10.1088/0264-9381/30/4/045001}{Class. Quant. Grav.
  {\bfseries 30}, 045001 (2013)}.

\bibitem{Zhang:2020smo}
X.~Zhang, J.~Yang, and Y.~Ma, {Canonical loop quantization of the lowest-order
  projectable Horava gravity},
  \href{http://dx.doi.org/10.1103/PhysRevD.102.124060}{Phys. Rev. D {\bfseries
  102}, 124060 (2020)}.

\bibitem{Ashtekar:2003hd}
A.~Ashtekar, M.~Bojowald, and J.~Lewandowski, {Mathematical structure of loop
  quantum cosmology}, \href{http://dx.doi.org/10.4310/ATMP.2003.v7.n2.a2}{Adv.
  Theor. Math. Phys. {\bfseries 7}, 233 (2003)}.

\bibitem{Ashtekar:2005qt}
A.~Ashtekar and M.~Bojowald, {Quantum geometry and the Schwarzschild
  singularity}, \href{http://dx.doi.org/10.1088/0264-9381/23/2/008}{Class.
  Quant. Grav. {\bfseries 23}, 391 (2006)}.

\bibitem{Bojowald:2005epg}
M.~Bojowald, {Loop quantum cosmology},
  \href{http://dx.doi.org/10.12942/lrr-2005-11}{Living Rev. Rel. {\bfseries 8},
  11 (2005)}.

\bibitem{Ashtekar:2006wn}
A.~Ashtekar, T.~Pawlowski, and P.~Singh, {Quantum nature of the big bang:
  Improved dynamics}, \href{http://dx.doi.org/10.1103/PhysRevD.74.084003}{Phys.
  Rev. D {\bfseries 74}, 084003 (2006)}.

\bibitem{Ashtekar:2011ni}
A.~Ashtekar and P.~Singh, {Loop quantum cosmology: A status report},
  \href{http://dx.doi.org/10.1088/0264-9381/28/21/213001}{Class. Quant. Grav.
  {\bfseries 28}, 213001 (2011)}.

\bibitem{Thiemann:1996aw}
T.~Thiemann, {Quantum spin dynamics (QSD)},
  \href{http://dx.doi.org/10.1088/0264-9381/15/4/011}{Class. Quant. Grav.
  {\bfseries 15}, 839 (1998)}.

\bibitem{Bojowald:2002gz}
M.~Bojowald, {Isotropic loop quantum cosmology},
  \href{http://dx.doi.org/10.1088/0264-9381/19/10/313}{Class. Quant. Grav.
  {\bfseries 19}, 2717 (2002)}.

\bibitem{Henriques:2006qb}
A.~B. Henriques, {Loop quantum cosmology and the Wheeler-De Witt equation},
  \href{http://dx.doi.org/10.1007/s10714-006-0330-1}{Gen. Rel. Grav. {\bfseries
  38}, 1645 (2006)}.

\bibitem{Yang:2009fp}
J.~Yang, Y.~Ding, and Y.~Ma, {Alternative quantization of the Hamiltonian in
  loop quantum cosmology},
  \href{http://dx.doi.org/10.1016/j.physletb.2009.10.072}{Phys. Lett. B
  {\bfseries 682}, 1 (2009)}.

\bibitem{Assanioussi:2018hee}
M.~Assanioussi, A.~Dapor, K.~Liegener, and T.~Paw\l{}owski, {Emergent de Sitter
  Epoch of the Quantum Cosmos from Loop Quantum Cosmology},
  \href{http://dx.doi.org/10.1103/PhysRevLett.121.081303}{Phys. Rev. Lett.
  {\bfseries 121}, 081303 (2018)}.

\bibitem{Li:2018opr}
B.-F. Li, P.~Singh, and A.~Wang, {Towards cosmological dynamics from loop
  quantum gravity}, \href{http://dx.doi.org/10.1103/PhysRevD.97.084029}{Phys.
  Rev. D {\bfseries 97}, 084029 (2018)}.

\bibitem{Zhang:2021zfp}
X.~Zhang, G.~Long, and Y.~Ma, {Loop quantum gravity and cosmological constant},
  \href{http://dx.doi.org/10.1016/j.physletb.2021.136770}{Phys. Lett. B
  {\bfseries 823}, 136770 (2021)}.

\bibitem{Vandersloot:2006ws}
K.~Vandersloot, {Loop quantum cosmology and the $k=-1$ Robertson-Walker model},
  \href{http://dx.doi.org/10.1103/PhysRevD.75.023523}{Phys. Rev. D {\bfseries
  75}, 023523 (2007)}.

\bibitem{Szulc:2007uk}
L.~Szulc, {An open FRW model in loop quantum cosmology},
  \href{http://dx.doi.org/10.1088/0264-9381/24/24/003}{Class. Quant. Grav.
  {\bfseries 24}, 6191 (2007)}.

\bibitem{Szulc:2006ep}
L.~Szulc, W.~Kaminski, and J.~Lewandowski, {Closed FRW model in loop quantum
  cosmology}, \href{http://dx.doi.org/10.1088/0264-9381/24/10/008}{Class.
  Quant. Grav. {\bfseries 24}, 2621 (2007)}.

\bibitem{Ashtekar:2006es}
A.~Ashtekar, T.~Pawlowski, P.~Singh, and K.~Vandersloot, {Loop quantum
  cosmology of $k=1$ FRW models},
  \href{http://dx.doi.org/10.1103/PhysRevD.75.024035}{Phys. Rev. D {\bfseries
  75}, 024035 (2007)}.

\bibitem{Mielczarek:2009kh}
J.~Mielczarek, O.~Hrycyna, and M.~Szydlowski, {Effective dynamics of the closed
  loop quantum cosmology},
  \href{http://dx.doi.org/10.1088/1475-7516/2009/11/014}{JCAP {\bfseries 11},
  014 (2009)}.

\bibitem{Corichi:2011pg}
A.~Corichi and A.~Karami, {Loop quantum cosmology of $k=1$ FRW: A tale of two
  bounces}, \href{http://dx.doi.org/10.1103/PhysRevD.84.044003}{Phys. Rev. D
  {\bfseries 84}, 044003 (2011)}.

\bibitem{Corichi:2013usa}
A.~Corichi and A.~Karami, {Loop quantum cosmology of $k=1$ FLRW: Effects of
  inverse volume corrections},
  \href{http://dx.doi.org/10.1088/0264-9381/31/3/035008}{Class. Quant. Grav.
  {\bfseries 31}, 035008 (2014)}.

\bibitem{Dupuy:2016upu}
J.~L. Dupuy and P.~Singh, {Implications of quantum ambiguities in $k=1$ loop
  quantum cosmology: Distinct quantum turnarounds and the super-Planckian
  regime}, \href{http://dx.doi.org/10.1103/PhysRevD.95.023510}{Phys. Rev. D
  {\bfseries 95}, 023510 (2017)}.

\bibitem{Barbero:1994ap}
J.~F. Barbero, {Real Ashtekar variables for Lorentzian signature space-times},
  \href{http://dx.doi.org/10.1103/PhysRevD.51.5507}{Phys. Rev. D {\bfseries
  51}, 5507 (1995)}.

\bibitem{Immirzi:1996dr}
G.~Immirzi, {Quantum gravity and Regge calculus},
  \href{http://dx.doi.org/10.1016/S0920-5632(97)00354-X}{Nucl. Phys. B - Proc.
  Suppl. {\bfseries 57}, 65 (1997)}.

\bibitem{Ding:2008tq}
Y.~Ding, Y.~Ma, and J.~Yang, {Effective scenario of loop quantum cosmology},
  \href{http://dx.doi.org/10.1103/PhysRevLett.102.051301}{Phys. Rev. Lett.
  {\bfseries 102}, 051301 (2009)}.

\bibitem{Ashtekar:2008zu}
A.~Ashtekar, {Loop quantum cosmology: An overview},
  \href{http://dx.doi.org/10.1007/s10714-009-0763-4}{Gen. Rel. Grav. {\bfseries
  41}, 707 (2009)}.

\bibitem{Ashtekar:2009um}
A.~Ashtekar and E.~Wilson-Ewing, {Loop quantum cosmology of Bianchi type II
  models}, \href{http://dx.doi.org/10.1103/PhysRevD.80.123532}{Phys. Rev. D
  {\bfseries 80}, 123532 (2009)}.

\bibitem{Assanioussi:2019iye}
M.~Assanioussi, A.~Dapor, K.~Liegener, and T.~Paw\l{}owski, {Emergent de Sitter
  epoch of the loop quantum cosmos: A detailed analysis},
  \href{http://dx.doi.org/10.1103/PhysRevD.100.084003}{Phys. Rev. D {\bfseries
  100}, 084003 (2019)}.

\bibitem{Zhang:2019dgi}
C.~Zhang, J.~Lewandowski, H.~Li, and Y.~Ma, {Bouncing evolution in a model of
  loop quantum gravity},
  \href{http://dx.doi.org/10.1103/PhysRevD.99.124012}{Phys. Rev. D {\bfseries
  99}, 124012 (2019)}.

\bibitem{Taveras:2008ke}
V.~Taveras, {Corrections to the Friedmann equations from loop quantum gravity
  for a universe with a free scalar field},
  \href{http://dx.doi.org/10.1103/PhysRevD.78.064072}{Phys. Rev. D {\bfseries
  78}, 064072 (2008)}.

\bibitem{Ashtekar:2010ve}
A.~Ashtekar, M.~Campiglia, and A.~Henderson, {Casting loop quantum cosmology in
  the spin foam paradigm},
  \href{http://dx.doi.org/10.1088/0264-9381/27/13/135020}{Class. Quant. Grav.
  {\bfseries 27}, 135020 (2010)}.

\bibitem{Qin:2012gaa}
L.~Qin, G.~Deng, and Y.~Ma, {Path integrals and alternative effective dynamics
  in loop quantum cosmology},
  \href{http://dx.doi.org/10.1088/0253-6102/57/2/28}{Commun. Theor. Phys.
  {\bfseries 57}, 326 (2012)}.

\bibitem{Liegener:2019out}
K.~Liegener and S.~A. Weigl, {Effective LQC model for $k$ = +1 isotropic
  cosmologies from spatial discretizations},
  \href{http://dx.doi.org/10.1103/PhysRevD.102.106014}{Phys. Rev. D {\bfseries
  102}, 106014 (2020)}.

\bibitem{Dapor:2017rwv}
A.~Dapor and K.~Liegener, {Cosmological effective Hamiltonian from full loop
  quantum gravity dynamics},
  \href{http://dx.doi.org/10.1016/j.physletb.2018.09.005}{Phys. Lett. B
  {\bfseries 785}, 506 (2018)}.

\bibitem{Han:2019vpw}
M.~Han and H.~Liu, {Effective dynamics from coherent state path integral of
  full loop quantum gravity},
  \href{http://dx.doi.org/10.1103/PhysRevD.101.046003}{Phys.\ Rev.\ D
  {\bfseries 101}, 046003 (2020)}.

\bibitem{Han:2019feb}
M.~Han and H.~Liu, {Improved $\bar{\mu}$-scheme effective dynamics of full loop
  quantum gravity}, \href{http://dx.doi.org/10.1103/PhysRevD.102.064061}{Phys.
  Rev. D {\bfseries 102}, 064061 (2020)}.

\bibitem{Liegener:2020dcg}
K.~Liegener and L.~Rudnicki, {Algorithmic approach to cosmological coherent
  state expectation values in loop quantum gravity},
  \href{http://dx.doi.org/10.1088/1361-6382/ac226f}{Class. Quant. Grav.
  {\bfseries 38}, 205001 (2021)}.

\bibitem{Zhang:2020mld}
C.~Zhang, S.~Song, and M.~Han, {First-order quantum correction in coherent
  state expectation value of loop-quantum-gravity Hamiltonian: Overview and
  results}, \href{http://arxiv.org/abs/2012.14242}{arXiv:2012.14242}.

\bibitem{Zhang:2021qul}
C.~Zhang, S.~Song, and M.~Han, {First-order quantum correction in coherent
  state expectation value of loop-quantum-gravity Hamiltonian},
  \href{http://dx.doi.org/10.1103/PhysRevD.105.064008}{Phys. Rev. D {\bfseries
  105}, 064008 (2022)}.

\bibitem{Liegener:2019zgw}
K.~Liegener and P.~Singh, {Gauge-invariant bounce from loop quantum gravity},
  \href{http://dx.doi.org/10.1088/1361-6382/ab7962}{Class.\ Quant.\ Grav.
  {\bfseries 37}, 085015 (2020)}.

\bibitem{Yang:2019ujs}
J.~Yang, C.~Zhang, and Y.~Ma, {Loop quantum cosmology from an alternative
  Hamiltonian}, \href{http://dx.doi.org/10.1103/PhysRevD.100.064026}{Phys. Rev.
  D {\bfseries 100}, 064026 (2019)}.

\bibitem{Yang:2020eby}
J.~Yang, C.~Zhang, and Y.~Ma, {Loop quantum cosmology from an alternative
  Hamiltonian. II. Including the Lorentzian term},
  \href{http://dx.doi.org/10.1103/PhysRevD.102.084018}{Phys. Rev. D {\bfseries
  102}, 084018 (2020)}.

\end{thebibliography}
\end{document}